\documentclass[a4paper,11pt]{article}
\pdfoutput=1 % if your are submitting a pdflatex (i.e. if you have
\usepackage{jheppub} % for details on the use of the package, please
\usepackage[T1]{fontenc} % if needed
\usepackage{xcolor}
\usepackage{bbm}
\usepackage{subfigure}
\usepackage{braket}
\usepackage{array}
    \newcolumntype{P}[1]{>{\centering\arraybackslash}p{#1}}
    \newcolumntype{M}[1]{>{\centering\arraybackslash}m{#1}}

\usepackage{hyperref}

\begin{document}

\title{\boldmath Symmetry-resolved entanglement entropy in Wess-Zumino-Witten models
}

\vspace{.5cm}

\author{Pasquale Calabrese$^{1,2}$, J\'er\^ome Dubail$^3$ and  Sara Murciano$^1$}
\affiliation{$^1$SISSA and INFN Sezione di Trieste, via Bonomea 265, 34136 Trieste, Italy.}
\affiliation{$^{2}$International Centre for Theoretical Physics (ICTP), Strada Costiera 11, 34151 Trieste, Italy.}
\affiliation{$^3$Universit\'e de Lorraine, CNRS, LPCT, F-54000 Nancy, France}
\emailAdd{smurcian@sissa.it}

\vspace{.5cm}

\abstract{We consider the problem of the decomposition of the R\'enyi  entanglement entropies in theories with a non-abelian symmetry by doing a thorough analysis of Wess-Zumino-Witten (WZW) models. We first consider $SU(2)_k$ as a case study and then generalise to an arbitrary non-abelian Lie group. We find that at leading order in the subsystem size $L$ the entanglement is equally distributed  among the different sectors labelled by the irreducible representation of the associated algebra. We also identify the leading term that breaks this equipartition: it does not depend on $L$ but only on the dimension of the representation. 
Moreover, a $\log\log L$ contribution to the R\'enyi entropies exhibits a universal prefactor equal to half the dimension of the Lie group.}% form related to the underlying symmetry group of the model, i.e. the dimension of the Lie group.} %Despite our results work for a generic non-abelian symmetry characterising the WZW-model, we consider $SU(2)_k$ as a case study.

\maketitle

\section{Introduction}

\paragraph{Entanglement entropy and quantum field theory.} 
Two-dimensional conformal field theories are characterised by an infinite-dimensional algebra, known as a Virasoro algebra, that leads to their exact solution \cite{belavin1984infinite,difrancesco,m-book}.
 There exists a set of field theories that present, in addition to conformal invariance, an internal Lie group symmetry: the Wess-Zumino-Witten models,
that possess interesting applications in a wide range of topics, such as the study of fundamental interactions, statistical mechanics, and condensed matter theory \cite{wzw2,wzw1,gnt-04,tsvelik}. 
In the present work, such theories will be the arena to study the decomposition of the entanglement into the charge sectors of the symmetry. \\
As widely known, when a system is in a pure state, the bipartite entanglement between a subsystem $A$ and its complementary $B$ may be quantified by the R\'enyi  entanglement entropies \cite{intro1,intro2,eisert-2010,intro3,h-19}
\begin{equation}
\label{eq:entr}
S_n=\frac{1}{1-n} \log\mathrm{Tr}_A \rho_A^{n},
\end{equation}
where $\rho_A$ is the reduced density matrix (RDM) of a subsystem $A$.
The von Neumann entropy is obtained as the limit $n \to 1$ of Eq. \eqref{eq:entr}, and the entire spectrum of $\rho_A$ can also be reconstructed from 
the R\'enyi entropies \cite{cl-08}. 
In field theory, the R\'enyi entropies $S_n$ are usually achieved through the replica approach because, for integer $n$, $\mathrm{Tr} \rho_A^{n}=\frac{Z_n}{Z_1^n}$ \cite{cc-04,cw-94} with
$Z_n$ the partition function on an $n$-sheeted Riemann surface obtained by joining cyclically the $n$ sheets along the region $A$. \\ 
This approach, when applied to the vacuum of a (1+1) dimensional conformal field theory (CFT), leads to the famous scaling results \cite{cc-04,cc-09,hlw-94,vidal,vidal1,cw-94}
\begin{equation}
\label{eq:entr2}
S_1(L)=\frac{c}{3} \log \frac{L}{\epsilon}, \qquad S_n(L)=\frac{c}{6}\frac{n+1}{n} \log \frac{L}{\epsilon} ,
\end{equation}
when the subsystem $A$ is an interval of length $L$ embedded in the infinte line and $\epsilon\ll L$ is an ultraviolet cutoff.
 \\ 
The possibility of measuring in an experiment the internal symmetry structure of the entanglement \cite{fis,vecd-20,vecd1-20,ahyrst-21} went together with 
new theoretical frameworks developed to address the same problem \cite{goldstein,xavier}. 
These progresses paved the way to study different symmetry-resolved contributions in various theoretical contexts such as CFTs 
\cite{goldstein, xavier,goldstein1,mbc-21,crc-20,c-21,cc-21,uv-21}, free \cite{mdgc-20,hcc-21} and interacting integrable quantum field theories \cite{dhc-20,hcc-21b},  
holographic settings \cite{bm-15,znm-20},  spin chains \cite{riccarda,SREE2dG,goldstein2,MDC-19-CTM,ccgm-20,lr-14,wv-03,bhd-18,bhd-19,bydm-20,mrc-20,bc-20,eimd-20,pbc-20,pbc-21,tr-19,vecd-20,vecd1-20,fg-21}, disordered systems \cite{trac-20,kusf-20,kusf-20b,kufs-20c} and for non-trivial topological phases \cite{clss-19,ms-20,as-20}.\\

\paragraph{Symmetry-resolved entanglement entropy.} The problem we address in this paper can be formulated as follows. We consider a 1+1D field theory on the infinite line $\mathbb{R}$, with the Hilbert space $\mathcal{H}$, and a symmetry group $G$ that acts unitarily on $\mathcal{H}$. The spatial bipartition $\mathbb{R} = A \cup B $ with $A = [0, L]$ and $B = (-\infty, 0) \cup (L, +\infty)$ corresponds to a bipartition of the Hilbert space of the field theory, $\mathcal{H} = \mathcal{H}_A\otimes \mathcal{H}_B$, and we assume that the action of $G$ is such that, for any element $g \in G$, the unitary matrix $U_g$ acting on $\mathcal{H}$ can be decomposed as $U_g = U^A_g \otimes U^B_g$, where $U^A_g$ ($U^B_g$) is a unitary operator acting on $\mathcal{H}_A$ ($\mathcal{H}_B$).

We focus on the symmetry-resolved entanglement entropy of the ground state of the quantum field theory. We assume that the ground state $\left| \psi_0 \right>$ is non-degenerate, so that it is invariant under the action of $G$: $U_g \left| \psi_0 \right> = \left| \psi_0 \right>$. Consequently, the reduced density matrix $\rho_A$ is also invariant (under the action of $G$ restricted to $\mathcal{H}_A$):
\begin{eqnarray}
\nonumber	U_g^A  \rho_A U_g^{A\dagger} &=& U_g^A  \left( {\rm Tr}_{B}  \left|  \psi_0 \right> \left< \psi_0 \right|  \right)  U_g^{A\dagger}   
%\\ \nonumber	&=&    
={\rm Tr}_{B}  \left( U_g^A  \left|  \psi_0 \right> \left< \psi_0 \right| U_g^{A\dagger}     \right)  \\
\nonumber	&=&    {\rm Tr}_{B}  \left( U_{g^{-1}}^B U_g  \left|  \psi_0 \right> \left< \psi_0 \right| U_g^{\dagger}  U_{g^{-1}}^{B \dagger}   \right)  
%\\\nonumber	&=&   
 = {\rm Tr}_{B}  \left( U_{g^{-1}}^B  \left|  \psi_0 \right> \left< \psi_0 \right|   U_{g^{-1}}^{B \dagger}   \right)  \\
	&=&    {\rm Tr}_{B}  \left|  \psi_0 \right> \left< \psi_0 \right|  \, = \, \rho_A ,
\end{eqnarray}
where we have used the cyclicity of the trace and the unitarity of $U_{g^{-1}}^B$ to arrive at the last line. 
Thus, when decomposing the Hilbert space $\mathcal{H}_A$ into a direct sum of irreducible representations of $G$,
the reduced density matrix $\rho_A$ is block-diagonal in the corresponding basis:
\begin{equation}\label{eq:probab}
	\rho_A  \,  = \, \bigoplus_r   p_r \rho_{A, r} \, = \,  \left( \begin{array}{c|c|c}
		p_{1} \rho_{A,1} &  \\  \hline
		 & p_{2} \rho_{A,2}  & \\ \hline
		  & & \ddots  \\
	\end{array} \right) .
\end{equation}
Here $r$ labels the irreducible representations of $G$.
Let us stress that if there are multiple occurrences of one representation $r$, they must be altogheter. 
For example, consider $G=SU(2)$, and a subsystem made of three spins $1/2$, whose decomposition into irreps is
\begin{equation}
(1/2) \otimes (1/2) \otimes (1/2)   =   2 (1/2) \oplus (3/2),
\end{equation}
where the spin (1/2) irrep appears with multiplicity $2$. In this example the density matrix would have two blocks, 
one with the two irreps of dimension $2$ transforming as the spin $1/2$-representation, the other with one irrep of dimension $4$
transforming as the spin $3/2$-representation under the action of $SU(2)$.
% In total there are only two representations corresponding to (3/2) and (1/2). 
%

In Eq. \eqref{eq:probab}, the block $\rho_{A,r}$ is normalised such that ${\rm Tr}\rho_{A,r} = 1$, and $p_r\equiv {\rm Tr} (\Pi_r \rho_A)$ is a non-negative number such that $ \sum_r p_r = {\rm Tr} \rho_A  = 1$ ($\Pi_r$ is the projector on the irrep $r$). 
%  a system which is invariant under the unitary action of a symmetry group $G$. The total Hilbert space, $\mathcal{H}$, is the tensor product $\mathcal{H}_A\otimes \mathcal{H}_B$ where $\mathcal{H}_A$ ($\mathcal{H}_B$) is a representation of the symmetry group acting on the subsystem $A$ ($B$). For such states we can build the RDM, $\rho_A$, which is also invariant under the action of the symmetry group acting on the subsystem $A$. This leads to a block diagonal structure of the reduced density matrix, where each block corresponds to an eigenspace of the generators of our symmetry (e.g. see Fig. \ref{fig:rdmimag}).
The {\it symmetry-resolved entanglement entropy} measures the entanglement in the subsystem $A$ for a fixed symmetry sector, i.e.
\begin{equation}\label{eq:RSREE}
	S_n^r   \, = \, \frac{1}{1-n } \log {\rm Tr}_A \rho_{A,r}^n , \qquad S_1^r =\lim_{n\to 1 }S_n^r.   
\end{equation}
The total von Neumann entanglement entropy  can be
written as \cite{nc-10,fis}
\begin{equation}\label{eq:dec}
S_1=\sum_r p_r S_1^r-\sum_r p_r\log p_r\equiv S^{\rm c}+S^{\rm n},
\end{equation}
where the first term, $S^{\rm c}$, is known as configurational entanglement entropy and measures the weighted sum of the entropy of all sectors, 
while the second one, $S^{\rm n}$, measures the fluctuation of the charge within $A$ and is called the number entropy.
The origin of the latter name is inspired by the case when the conserved charge is a particle number. 
Clearly such a name is inappropriate for a more general charge like the ones considered here, but we prefer not to change a well established terminology. 
It is worth mentioning that the relation in Eq. \eqref{eq:dec} has also been exploited to study the entanglement structure for
both abelian and non-abelian (lattice) gauge theories, e.g. in \cite{bccjk-17,gst-15,st-15,abhmsv-16,chr-14,aity-17}.
In the gauge theories, the symmetry resolved entropy $S_1^r$ is further split into two contributions.

\paragraph{Entanglement equipartition.} One important finding in the aforementioned literature is that conformal invariance forces the entanglement entropy to be equally distributed among the different sectors of a $U(1)$ symmetric theory \cite{xavier}. 
As an example, we can consider a massless compact boson (aka Luttinger liquid), which is a CFT with central charge $c = 1$ and
a $U(1)$ symmetry generated by the current operator $J(z) = i\sqrt{K}\partial \phi(z)$, where $\phi(z)$ is a chiral boson and $K$ a constant related to the compactification radius of the theory. 
The entanglement entropy in the sector labelled by the charge eigenvalue $q$ has been found to be \cite{goldstein,xavier,riccarda}
\begin{equation}\label{eq:intro}
S_{n}^{q}(L)=S_{n}(L)-\frac{1}{2}\log \Big( \frac{2K}{\pi}\log L\Big) +\frac{1}{2} \frac{\log(n)}{1-n}+o(L^0) .
\end{equation}
It may look awkward that the symmetry-resolved contributions have a double log correction, while the total entanglement entropy does not. However, when calculating the latter according to Eq. \eqref{eq:dec} this double log cancels \cite{MDC-19-CTM,riccarda}. 

The independence of Eq. \eqref{eq:intro} on the charge $q$ has been dubbed entanglement {\it equipartition} \cite{xavier},
which is the main feature of a CFT endowed with an abelian symmetry. It is an open issue to understand whether such equipartition of entanglement survives for a CFT with a non-abelian symmetry. To date, there are no results concerning such theories, with the exception of the $SU(2)$ spin-chain/CFT studied in \cite{goldstein}.  
However, the authors used different conventions with respect to ours, since they do not normalise each block of $\rho_A$ by each trace before calculating the entropies, so the resulting resolved entropies are not entanglement measures by themselves. 
Another important aspect of Eq. \eqref{eq:intro} is that the dependence on the ultraviolet cutoff $\epsilon$ at order $O(L^0)$ is fully encoded in the total entropy 
and so, Eq. \eqref{eq:intro} is universal. Equivalently, we can say that $S_n(L)-S_n^q(L)$ is universal up to order $L^0$.
It is also natural to wonder whether such universality persists for non-abelian symmetries.  

\paragraph{1+1D CFTs with non-abelian symmetry: Wess-Zumino-Witten models.} Our goal is to fill this gap and to study how the total entanglement splits into the contributions coming from symmetry sectors in CFTs with a non-abelian Lie group symmetry, i.e. the Wess-Zumino-Witten (WZW) models. They are described by a two-dimensional action which consists of a non-linear $\sigma$ term plus the Wess-Zumino term, whose topological coupling factor $k$ is constrained to be an integer number and it is referred to as the level of the model \cite{wzw1,wzw2}. 
Here we follow the conventions of Ref.~\cite{difrancesco}. For simplicity, we assume that the Lie group $G$ is compact and simple. \\
These WZW models are the scaling limit of critical quantum spin chains with the same symmetry \cite{tsvelik,f-13,gnt-04}. For instance, some possible discretisations of $SU (2)_k$ are the Heisenberg spin-1/2 chain or the Haldane-Shastry model for $k=1$  \cite{aff1,aff2,ah-87,h1,h2}. They have been also studied in the context of topological anyons on 1D chains \cite{gath-13}. Spin chains associated with spin $j=k/2$, $k>1$ correspond to $SU(2)_k$ WZW models, which can mark phase transitions between different gapped phases, as in the Babudjan-Takhtajan chains \cite{b1,b2,bondesan2015chiral}.

\paragraph{Main result.} In this paper we compute the symmetry-resolved entanglement entropy in the ground state of a non-abelian WZW model. 
In the course of the derivation we make some assumptions on the multiplicity of the entanglement spectrum, or equivalently on the 
conformal boundary conditions induced at the entangling points (see Sections \ref{sec:4}, \ref{sec:5} and \cite{ct}). With these assumptions, we
find that, for large $L$, 
\begin{eqnarray}\label{eq:SREE_final}
\nonumber S_n^r (L)  & {=}&  S_n(L) -  \frac{{\rm dim} (G)}{2} \log (\log L)  \\
\nonumber	&&  +2 \log {\rm dim}(r)  - \log   \frac{{\rm Vol}(G)}{|Z(G)| }   + \frac{{\rm dim} (G)}{2} \left( - \log k + \frac{\log n}{1-n} + \log( 2\pi^3 ) \right)  \\  
	&&  + o(L^0) ,
\end{eqnarray}
where ${\rm dim} (G)$ is the dimension of the Lie group $G$, $\mathrm{Vol}(G)$ is its volume, and $\mathrm{dim}(r)$ is the dimension of the representation. Here $Z(G)$ is the center of $G$, which is a finite subgroup whose order is denoted by $|Z(G)|$. This result extends the abelian one, see Eq. \eqref{eq:intro}, since there $\dim r=1$ for all sectors.  
However, in contrast with the latter case, now the entanglement explicitly depends on the charge sectors at ${O}(L^0)$. 
It is also consistent with the one of Ref.~\cite{goldstein} for the $SU(2)$ case. %see section~\ref{sec:2}.
The only difference is that here we are interested in a symmetry decomposition of $SU(2)$ with respect to the total spin, 
 while Ref.~\cite{goldstein} reports the resolution with respect to both the total spin and its $z$-component, see section~\ref{sec:2}.
Our approach also provides an explicit expression for $p_r$ (see Eq. \eqref{eq:probab11}). 
We mention that different assumptions on the entanglement spectrum would change, in a calculable way, the $O(L^0)$ terms in Eq. \eqref{eq:SREE_final}, but 
leave unchanged the double logarithm whose prefactor depends only on the dimension of the group.
Finally we mention that the ${\rm dim}(r)$ dependence in Eq. \eqref{eq:SREE_final} is reminiscent of the representation entropy introduced in the context of 
gauge theories \cite{bccjk-17,gst-15,st-15,abhmsv-16,chr-14}.

\paragraph{Outline.} The paper is structured as follows. In section \ref{sec:2}, we provide all the definitions concerning the measures of symmetry-resolved entanglement and we review the example of a non-abelian resolution for a $SU(2)$ spin chain. In section \ref{sec:3}, we present the WZW-models, its symmetry algebra and we introduce the notion of character of a representation.  
Using the modular properties of unspecialised characters, we calculate the moments of the RDM in presence of a charge flux, that we call charged moments and we give an alternative derivation of the symmetry decomposition of entanglement for WZW-models with $SU(2)_k$ symmetry in section \ref{sec:4}. 
This strategy has the advantage of being generalisable for the computation of the symmetry-resolved entanglement entropies for an arbitrary non-abelian symmetry, 
as showed in section \ref{sec:5}. We conclude in section \ref{sec:concl}. 
Two appendices are also included: in Appendix  \ref{app:app1}, we review how the RDM can be expressed in terms of Virasoro generators while in Appendix \ref{app:app2} 
we give an example of a field theory with a non-abelian symmetry but which is not a WZW-model.

\section{Overview of known results for $U(1)$ and $SU(2)$}\label{sec:2}
In this section we review the known results about the symmetry resolution of an abelian symmetry and a $SU(2)$ one, following Ref.~\cite{goldstein}. 
Reviewing the $U(1)$ case allows us to introduce some of the key steps that we will adapt to non-abelian Lie groups $G$ in Sections~\ref{sec:4} and~\ref{sec:5}. 
In contrast, the method of Ref.~\cite{goldstein} for the $SU(2)$ case is not generalisable to other groups $G$; 
nevertheless we review it here for completeness and as a comparison for our main derivation.
The more general method is presented in details in Sections~\ref{sec:4} and~\ref{sec:5}.

\subsection{$U(1)$ symmetry-resolved entanglement (after Ref.~\cite{goldstein})}\label{sec:U1}
We consider a system with internal $U(1)$ symmetry, generated by a charge operator $Q$ with eigenvalues in $\mathbb{Z}$: the group elements $e^{i \alpha} \in U(1)$ act on the Hilbert space $\mathcal{H}$ as $e^{i \alpha Q}$.
For a bipartition into two subsystems, $A$ and $B$, the charge $Q$ is the sum of the charges in $A$ and $B$, $Q = Q_A+Q_B$. The reduced density matrix $\rho_A$ admits a decomposition according to the eigenvalue $q \in \mathbb{Z}$ of the charge operator $Q_A$, 
\begin{equation}
\label{eq:sum}
\rho_A=\bigoplus_q p_q \rho_{A,q},
\end{equation}
where $p_q$ is the probability of finding $q$ in a measurement of $Q_A$ in the RDM $\rho_A$, i.e. $p_q= \mathrm{Tr} (\Pi_q \rho_A)$ and $\Pi_q$ is the projector onto the subspace of eigenvalue $q$. 
The density matrices $\rho_{A,q}$ of different blocks are normalised: ${\rm Tr}\rho_{A, q}=1$.  
The symmetry-resolved entropies (\ref{eq:RSREE}) can be obtained  from the entanglement spectrum of $\rho_A$ and its resolution in the charge sectors, see Appendix \ref{app:app1}. Another way to compute them is to introduce the {\it charged moments} of the reduced density matrix as in Ref.~\cite{goldstein},
\begin{equation}
\label{eq:firstdef}
	Z_n(\alpha,L)\equiv	Z_1^n \, \mathrm{Tr}_A \left[ \rho_A^n \,e^{i \alpha Q_A } \right],
\end{equation}
where we keep the factor $Z_1^n$ to ensure that for $\alpha=0, n=1$, $\mathrm{Tr}_A\rho_A=1$.
Using the Fourier representation of the projection operator, we get the moments of the RDM restricted to the sector of fixed charge $q$ \cite{goldstein},
\begin{equation}
\label{eq:defF}
\mathcal{Z}_n^q(L)\equiv \mathrm{Tr} (\Pi_{q}\,\rho^n_A)=\displaystyle \int_{-\pi}^{\pi}\dfrac{d\alpha}{2\pi}e^{-iq\alpha} \mathrm{Tr}_A \left[ \rho_A^n \,e^{i \alpha Q_A } \right].
\end{equation}
The probability introduced in Eq. \eqref{eq:probab} is $p_q(L)=\mathcal{Z}_1^q(L)$.
Finally, the symmetry-resolved entropies are obtained as
\begin{equation}
\label{eq:SREE1}
S_n^q(L)=\dfrac{1}{1-n}\log \left[ \dfrac{\mathcal{Z}^q_n(L)}{(\mathcal{Z}^q _1(L))^n}\right] . %\qquad S_{1}^q(L)=\lim_{n\rightarrow 1} S^q_n(L).
\end{equation}
This formalism is easily applied to the a free massless compact boson, which is the easiest CFT with $U(1)$ symmetry. 
The charged moments for a single interval of length $L$ on the infinite line behave as \cite{goldstein} 
\begin{equation}\label{eq:chargedmom}
	\frac{Z_n(\alpha,L)}{Z_1^n}  =  c_{n,\alpha}L^{- \frac{c}{6} (n -\frac{1}{n}) - \frac{2K}{n}(\frac{\alpha}{2\pi})^2},  
\end{equation}
where $c_{n,\alpha}$ is a non-universal constant which depends on the cutoff (i.e. the microscopic details of the model).
Eq. \eqref{eq:chargedmom} is valid for $\alpha \in (-\pi,\pi)$, but what we really need is just its behaviour around $\alpha=0$. Indeed, we can evaluate the Fourier transform (\ref{eq:defF}) by the saddle point approximation at large $L$ to get
\begin{multline}
	\label{eq:ressaddleU1}
\mathcal{Z}_n^q(L) \, = \,  L^{- \frac{c}{6} (n -\frac{1}{n})  } \displaystyle \int \dfrac{d\alpha}{2\pi}e^{-iq\alpha}  e^{- \frac{2K}{n}(\frac{\alpha}{2\pi})^2  \log L}  c_{n,\alpha}\\
\simeq c_{n,0}L^{- \frac{c}{6} (n -\frac{1}{n})  }  \frac{(\pi n)^{1/2} }{(2K)^{1/2} (\log L)^{1/2}}  e^{-\frac{nq^2\pi^2}{2K\log(L/\epsilon)}}.
\end{multline}
Plugging this into Eq. (\ref{eq:SREE1}) leads to the symmetry-resolved R\'enyi entropy~(\ref{eq:intro}). 
The effect of the $\alpha$ dependence of $c_{n,\alpha}$ is to renormalise the variance of the distribution in Eq. \eqref{eq:ressaddleU1} by a $O(L^0)$ term and so 
leading to a correction $O(1/(\log L))$ terms in the entropy \eqref{eq:intro}, as shown for a free fermionic system in \cite{riccarda,eimd-20}.

\subsection{$SU(2)$ symmetry-resolved entanglement (after Ref.~\cite{goldstein})}\label{sec:su2g}

 To tackle the $SU(2)$ case, the authors of Ref.~\cite{goldstein} rely on the following trick, which allows them to recycle the result for the $U(1)$ case. Decomposing the Hilbert space $\mathcal{H}_A$ into $SU(2)$ sectors with spin $j$ and magnetisation $j^z$, they notice that, for an $SU(2)$-invariant reduced density matrix $\rho_A$, 
\begin{equation}
\label{eq:coeff4}
\mathrm{Tr}_{J_A=j}\rho_A^n=(2j+1)(\mathrm{Tr}_{J_A^z=j}\rho_A^n-\mathrm{Tr}_{J_A^z=j+1}\rho_A^n).
\end{equation}
Here the trace in the left-hand side is over all states in $\mathcal{H}_A$ with spin $J_A= j$, while the two traces in the right-hand side are over all states with fixed magnetisation  $J_A^z$, without  restriction on the total spin $J_A$. Eq. \eqref{eq:coeff4} is slightly different from the identity used in \cite{goldstein}:
\begin{equation}\label{eq:fixjz}
{\rm Tr}_{J_A=j,J_A^z=j^z}=\mathrm{Tr}_{J_A^z=j}\rho_A^n-\mathrm{Tr}_{J_A^z=j+1}\rho_A^n,
\end{equation} 
i.e. the trace in the left-hand side is over all states with fixed spin $J_A= j$ and $J_A^z=j^z$, whose multiplet structure gives the contribution $(2j+1)$ in our Eq. \eqref{eq:coeff4}.
 A sketch of a derivation of that identity is given below. First, let us explain how the $SU(2)$ symmetry-resolved entanglement can be obtained from there. The point is that the operator $J^z_A$ generates a $U(1)$ symmetry (an abelian subgroup of $SU(2)$) so it can be identified with the charge operator $Q$ of the previous section (up to an unimportant constant). 
The two terms in the right-hand side of Eq.\,(\ref{eq:coeff4}) can be computed using the same methods as for the $U(1)$ case. 
Indeed, the charged moments related to the $U(1)$ subgroup are just given by Eq. \eqref{eq:chargedmom}, with $K=k/2$, see Refs. \cite{goldstein,xavier} and section \ref{sec:charged}. 
%As we will explain in detail in section \ref{sec:charged}, the charged moments for  the $SU(2)_k$ WZW model behave as
%\begin{equation}
%	 \frac{Z_n(\alpha,L)}{Z_1^n} =c_{n,\alpha} L^{ - \frac{c}{6} (n-\frac{1}{n}) -\frac{k}{n}(\frac{\alpha}{2\pi})^2}
%%Z_n(\alpha,L)=\frac{Z_n}{Z_1^n}L^{-\frac{k}{n}(\frac{\alpha}{2\pi})^2},
%\end{equation}
%at large $L$. 
%In the difference in the right hand side of Eq. \eqref{eq:coeff4} the leading order in Eq.~(\ref{eq:ressaddleU1}) cancels, so we need to evaluate 
The saddle point approximation of the Fourier transform gives
\begin{eqnarray}
\mathcal{Z}_n^{J^z_A=j}(L) &=&  \frac{1}{Z_1^n}\int_{-\pi}^{\pi} \frac{d \alpha}{2\pi} e^{- i \alpha j}   Z_n(\alpha,L) \\
\nonumber  & \simeq & c_{n,0}     L^{- \frac{c}{6} (n - \frac{1}{n})} e^{-\frac{nj^2\pi^2}{k\log(L/\epsilon)}} \left( \frac{(\pi n)^{1/2}}{k^{1/2} (\log L)^{1/2}} +\dots\right) .
\end{eqnarray}
where the dots stands for neglected subleading contributions, due e.g. to $c_{n,\alpha}$.
From Eq.\,(\ref{eq:coeff4}) we get 
\begin{equation}\label{eq:FG}
\mathcal{Z}_n^{J_A=j} (L)  \simeq c_{n,0} e^{-\frac{nj^2\pi^2}{k\log(L/\epsilon)}} L^{- \frac{c}{6} (n - \frac{1}{n})} (2 j +1)^2  \frac{ \pi^{5/2}n^{3/2}}{k^{3/2} (\log L)^{3/2}} ,
\end{equation} 
where we keep the Gaussian factor in order to have a normalised probability $\mathcal{Z}_1^{J_A=j} (L) $.
Eq. \eqref{eq:FG} leads to the desired symmetry-resolved R\'enyi entropy for the spin-$j$ representation:
\begin{multline}\label{eq:car1}
%S_{n}^{j}(L)&=S_n(L)  -\frac{3}{2}\log(k \log L) +\log( 2 j+1)+\frac{3\log (n)}{2(1-n)}+\frac{5}{2}\log (\pi)+o(L^0).
S_{n}^{j}(L)=
S_n(L)  -\frac{3}{2}\log( \log L) \\+2\log( 2 j+1) - \log (2^{3/2} \pi^2) +   \frac{3}{2} \left( - \log k +  \frac{\log (n)}{1-n} + \log (2 \pi^3)  \right)  +o(L^0) .
%S_{1,k}^{J_A,J^z_A}(L)&=S_{1}(L)+\log ( 2j+1)-\frac{3}{2}-\frac{3}{2}\log(k \log L)+o(L^0).
\end{multline} 
This result is the first and unique example of a non-abelian symmetry resolution in the literature, but it cannot be easily generalised to an arbitrary non-abelian symmetries 
because it strongly relies on the identity (\ref{eq:coeff4}). However, an alternative derivation of the same result in Section \ref{sec:4} will allow us to perform such generalisation.

\paragraph{Derivation of the identity (\ref{eq:coeff4}).} Now let us come back to the identity (\ref{eq:coeff4}), on which the calculation of Ref.~\cite{goldstein} relies. Because the matrix $\rho_A^n$ is invariant under the action of $SU(2)$, it can be decomposed as a sum of projectors over all irreducible representations appearing in the decomposition of $\mathcal{H}_A$ (here each irrep of spin $j$ can appear several times, which is accounted for by the index $a$):  
\begin{equation}
	\rho_A^n = \sum_j \sum_a  c_{(j,a)}  \Pi_{(j,a)} ,
\end{equation}
with some coefficients $c_{(j,a)}$. We have $ \mathrm{Tr} \,  \Pi_{(j,a)}   =   2j+1$ and, with the same notations as in Eq.~(\ref{eq:coeff4}), $ \mathrm{Tr}_{J^z_A=j}  \Pi_{(j,a)} = 1$. More generally,
\begin{equation}
	\mathrm{Tr}_{J_A^z=j}  \Pi_{(j',a)}  =  \left\{ \begin{array}{rcl}
		1 && {\rm if} \quad j \leq j' \\
		0  && {\rm otherwise} ,
		\end{array} \right.
\end{equation}
because the irrep of spin $j'$ contains exactly one state with $j^z = j$ if $j \leq j'$, and zero otherwise. Consequently,
\begin{equation}\label{eq:j1}
	\mathrm{Tr}_{J^z_A=j}  \Pi_{(j',a)}  - \mathrm{Tr}_{J^z_A=j+1}  \Pi_{(j',a)}  \, = \, \delta_{j,j'} ,
\end{equation}
which gives Eq.~(\ref{eq:coeff4}). 
Since Eq. \eqref{eq:j1} works for fixed $j^z$, if we do not consider any restriction on the values of $j^z$, we get the result in Eq. \eqref{eq:coeff4}, otherwise we obtain Eq. \eqref{eq:fixjz}.

\section{WZW models: currents and characters}\label{sec:3}
In this section we introduce our conventions for WZW models and review some fundamental objects which will be  useful later on. The interested readers can consult the comprehensive literature on the subject, for example in \cite{difrancesco,kac,wzw2,m-book}.

\subsection{WZW model on $G$ and current algebra}

We consider a compact simple Lie group $G$ and the associated Lie algebra $\mathfrak{g} = {\rm Lie}(G)$. Let $J^{a}$ ($a=1,\dots, {\rm dim}~\mathfrak{g}$) be generators of $\mathfrak{g}$, with commutation relations
\begin{equation}
	[ J^a, J^b ] \,= \,\sum_c i f^{ab}_{\phantom{aa} c} J^c ,
\end{equation}
with structure constants $f^{ab}_{\phantom{aa} c}$.\\
In the WZW model on the Lie group $G$, the symmetry is locally generated by the holomorphic and anti-holomorphic current $J^a (z)$ and $\overline{J}^a (\bar{z})$, where $(z,\bar{z})$ are complex coordinates for 2D Euclidean space. As usual in CFT, the holomorphic and anti-holomorphic components are independent and isomorphic. Focusing on the holomorphic components, the modes in their Laurent expansion
\begin{equation}\label{eq:modes}
J^{a}(z)=\sum_{n=-\infty}^{\infty}\frac{J_n^a}{z^{n+1}}
\end{equation}
obey the commutation relations of the Kac-Moody algebra at level $k$ (the level $k$ is a positive integer),
\begin{equation}\label{eq:algebra}
[J^{a}_n,J^{b}_m]= i \sum_c f^{ab}_{\phantom{aa}c}J^c_{m+n} +k \, m  \,  K(J^a, J^b) \, \delta_{m+n,0}.
\end{equation}
Here,
\begin{equation}\label{eq:killing}
	K(X, Y)  \equiv \frac{1}{2 \textsl{g}} \mathrm{Tr}({\rm ad} X {\rm ad Y})
\end{equation}
is the Killing form of $\mathfrak{g}$, which 
is positive definite because $G$ is compact. We follow the normalisation convention of Ref.~\cite{difrancesco}, with the inclusion of the factor $\frac{1}{2 \textsl{g}}$ where $\textsl{g}$ is the dual Coxeter number of $\mathfrak{g}$. \\
The currents can be multiplied to construct the energy-momentum tensor, whose mode expansion generates the Virasoro algebra. Mathematically, this means that the enveloping algebra of the Kac-Moody algebra contains a subalgebra that is the Virasoro algebra, a result which is known as the Sugawara construction \cite{difrancesco}. We can express the energy-momentum tensor in terms of the currents in the following way:
\begin{equation}\label{eq:T}
T(z)=\frac{1}{2(k+\textsl{g})}\sum_a :J^a(z) J^a(z):,
\end{equation}
where $::$ denotes the normal ordering, which consists in the subtraction of the singular terms. The computation  of the operator product expansion $T(z)T(w)$ determines the central charge $c$ of the theory, which is
\begin{equation}\label{eq:cc}
c= \frac{k~{\rm dim}(G)}{k+\textsl{g}},
\end{equation}
where ${\rm dim}(G) = {\rm dim}~\mathfrak{g}$ is the dimension of the Lie group $G$, or equivalently the dimension of $\mathfrak{g}$. As already mentioned, the stress-energy tensor can be expanded into mode operators, $L_n$, the Virasoro generators, that read 
\begin{equation}\label{eq:virsug}
L_n=\frac{1}{2\pi i}\oint dz\,z^{n+1}T(z)=\frac{1}{2(k+\textsl{g})}\sum_a\sum_m :J_m^{a }J^a_{n-m}:,
\end{equation}
where the integration contour circles the origin and the normal ordering means that positive modes should appear to the right of negative ones.

\subsection{The unspecialised characters and their asymptotic behaviour}\label{sub:3.1}

Primary fields of WZW models are in one-to-one correspondence with highest weight representations of the Kac-Moody algebra (\ref{eq:algebra}), see e.g. Ref.~\cite{difrancesco}. Each primary field transforms as a representation $r$ under conjugation by elements of $G$, so we can label them by irreps of $G$. The {\it unspecialised character} of the corresponding highest weight representation of the Kac-Moody algebra $\mathcal{M}_r$ is defined as 
\begin{equation}\label{eq:unspecialised}
\chi_r({\bf x},\tau)=\mathrm{Tr}_{\mathcal{M}_{r}}e^{ i \sum_{a}x_{a}J_0^{a}}e^{2\pi i \tau(L_0-\frac{c}{24})}.
\end{equation}
Here ${\bf x} = (x_1,\dots, x_{{\rm dim }~\mathfrak{g}})$ is the coordinate of elements in the Lie algebra, so $e^{ i \sum_{a}x_{a}J^{a}}$ is viewed as an element of $G$ via the exponential map. When ${\bf x}=0$, Eq. \eqref{eq:unspecialised} is referred to as the {\it specialised character}, $\chi_r(\tau) := \chi_r(0,\tau)$. We note that, in the literature (see e.g.~\cite{kac}), 
unspecialised characters are sometimes defined alternatively as
\begin{equation}
\mathrm{Tr}_{\mathcal{M}_{r}}e^{  i \sum_{b}\alpha_{b} H_0^{b}}e^{2\pi i \tau(L_0-\frac{c}{24})}, 
\end{equation}
where $H^b$ ($b=1,\dots , {\rm rank}~\mathfrak{g}$) are Cartan generators (i.e. the generators of a maximal commuting subalgebra of $\mathfrak{g}$), so that $\sum_{b}\alpha_{b} H^{b}$ is an element of the Cartan subalgebra $\mathfrak{h} \subset \mathfrak{g}$, instead of an arbitrary element of $\mathfrak{g}$. We stress that this makes no difference, because any element of $\mathfrak{g}$ is conjugated to an element of $\mathfrak{h}$. In other words, for any element $\sum_{a}z_{a}J^{a} \in \mathfrak{g} $, there exists $g \in G$ such that $\sum_{a}z_{a}J^{a} = g^{-1} \sum_{b}\alpha_{b} H^{b} g$ for some $\sum_{b}\alpha_{b} H^{b} \in \mathfrak{h}$. Using the fact $[g,L_0] = 0$ and the cyclicity of the trace, one sees that the two definitions are equivalent.\\
In what follows, we will need the asymptotics of $\chi_r({\bf x},\tau)$ when $\tau \to i0^+$. This is obtained by using the modular properties of the characters. Under the modular transformation $\tau \to -1/\tau$, the unspecialised character \eqref{eq:unspecialised} transforms as  
 \cite{kac,difrancesco}
\begin{equation}\label{eq:modular}
\chi_r({\bf x} ,\tau)=e^{-i \frac{k }{4\pi \tau} K ( {\bf x} \cdot {\bf J}  , {\bf x} \cdot {\bf J} )} \sum_{r'}S_{rr'}~\chi_{r'}\left(\frac{{\bf x}}{\tau},-\frac{1}{\tau}\right),
\end{equation}
where $S$ is the modular $S$-matrix, which is unitary and symmetric. In the argument of the exponential, we use the notation ${\bf x} \cdot {\bf J}  = \sum_a x_a J^a $, and $K(.,.)$ is the positive definite Killing form, normalised as in Eq.~(\ref{eq:killing}).
As explained in the following sections, we are mainly interested in the behaviour of characters around the elements of the center of $G$, e.g. the unit element. Therefore, in the limit $\tau \to i0^+$, we can keep only the leading contribution of each character $\chi_{r'}\left(-\frac{1}{\tau}\right)$, i.e.
\begin{equation}\label{eq:asymp1}
\chi_r({\bf x} ,\tau)\underset{{\tau \to i0^+}}{\simeq} e^{-i \frac{ k }{4\pi \tau} K( {\bf x} \cdot {\bf J} , {\bf x} \cdot {\bf J}  ) } \sum_{r'}S_{rr'}e^{-\frac{2\pi i }{\tau}(h_{r'}-\frac{c}{24})},
\end{equation}
where $h_{r'}$ is the conformal dimension of the primary field associated with the highest weight representation $r'$. The leading behaviour of Eq. \eqref{eq:asymp1} is given by the smallest dimension field. Since we are dealing with unitary theories, this is given by the identity with $h_{0}=0$, while the conformal dimensions of all other fields are  strictly positive. As a consequence, we have \cite{difrancesco}
\begin{equation}\label{eq:asymptotics}
\chi_r({\bf x},\tau) \underset{{\tau \to i0^+}}{\simeq} S_{r,0}~e^{\frac{\pi i c }{12 \tau}}~e^{-i \frac{ k }{4\pi \tau} K( {\bf x} \cdot {\bf J},  {\bf x} \cdot {\bf J} ) } .
\end{equation}
This asymptotic behaviour plays a key role in our derivation of the symmetry-resolved entanglement entropy below.

\subsection{Haar measure on $G$ from the Killing form, and orthonormality of group characters}

\paragraph{Metric and Haar measure.} Importantly, because $G$ is compact and simple, the Killing form $K(.,.)$ is positive definite on $\mathfrak{g}$. The Killing form then gives rise to a Riemannian metric on $G$. In a local coordinate chart ${\bf x} \in \mathbb{R}^{{\rm dim}(G)} \mapsto  g({\bf x}) \in G$, this metric can be defined as follows:
\begin{equation}\label{eq:Kab}
	K_{a b} ({\bf x}) \equiv K( i~g^{-1}({\bf x}) \partial_a g({\bf x}) , i~g^{-1}({\bf x}) \partial_b g({\bf x})  ) .
\end{equation}
[The factors $i$ come from the fact that we use the physics convention that the Lie algebra elements $X \in \mathfrak{g}$ are multiplied by $i$ before being exponentiated to give a group element $e^{i X} \in G$. Then $  g^{-1}({\bf x}) \partial_b g({\bf x})$ needs to be multiplied by $i$ to be in the Lie algebra.]
This metric induces a volume form on $G$,
\begin{equation}
	\label{eq:killing_haar}
	d \mu (g({\bf x}) ) \equiv  \sqrt{ {\rm det} K({\bf x})} \, d{\bf x} ,
\end{equation}
which turns out to be the Haar measure on $G$. We recall that the Haar measure is unique up to normalisation, and here the normalisation of the measure $d \mu$ is fixed by the normalisation of the Killing form. In particular, the volume of the group
\begin{equation}
	{\rm Vol}(G) \equiv  \int_G d \mu (g) 
\end{equation}
is fixed by this normalisation convention \cite{mc}.

To see that the measure (\ref{eq:killing_haar}) is the Haar measure on $G$, one can check that it is invariant under left multiplication by a fixed group element $h \in G$,
\begin{equation}
	d \mu (h g) =  d \mu (g) , 
\end{equation}
which is a consequence of the invariance of the Killing form under conjugation by elements of $G$,
\begin{equation}
	K( h^{-1} X h ,  h^{-1} Y h ) \, = \, K(X,Y),  \qquad {\rm for \; all} \quad h \in G.
\end{equation}
Moreover, for compact Lie groups, a left invariant measure must also be right invariant, i.e. $d \mu (g h) =  d \mu (g)$, so it is the Haar measure on $G$.

\paragraph{Group characters.} 
Finally, let us recall the definition of group characters.
 Given $g \in G$, a representation $U_g$ is not unambiguous since any similarity transformation yields an equivalent form. In order to describe the invariant properties of the group, one could use the eigenvalues of a representation matrix which do not change under similarity transformations. This leads to the construction of the {\it Casimir operators}, the eigenvalues of which classify the representation. Since this is in general a very difficult problem, in many cases it is sufficient to use a simpler invariant, namely the group character of a representation $r$, which is defined in terms of the unitary matrix $U_g$ as 
 \begin{equation}
 \chi_r(g)=\mathrm{Tr}U_g,
 \end{equation}
and it is invariant under similarity transformations. Importantly, group characters of irredubile representations are orthonormal with respect to the Haar measure,
\begin{equation}
\label{eq:orthogonality}
\frac{1}{{\rm Vol}(G)} \int_G d\mu (g) \chi_r(g) \chi^*_{r'}(g) =\delta_{rr'} .
\end{equation}

\section{Revisiting the $SU(2)_k$ case}\label{sec:4}
In this section we provide a detailed derivation of the charged moments for $SU(2)_k$ and the corresponding entanglement decomposition. This alternative approach with respect to the one reviewed in section \ref{sec:su2g} leads to a generalisation to an arbitrary non-abelian symmetry reported in the next section.

\subsection{The entanglement Hamiltonian}

We focus on a critical system described by the WZW-model $SU(2)_k$ at level $k$, and central charge $c=\frac{3k}{k+2}$, where we used Eq. \eqref{eq:cc} with $\textsl{g}=2$. 
In CFT, the powers of the reduced density matrix, $\rho_A^{n}$, are expressed as \cite{hlw-94,ct}
\begin{equation}\label{eq:bw1m}
\rho_A^n=\frac{e^{-2\pi n K_A}}{Z_1^n}, \quad Z_1=\mathrm{Tr}_A e^{-2\pi  K_A},
\end{equation}
where $K_A =\int_A dx T_{00}(x)/f'(x)$ is the entanglement Hamiltonian and $T_{00}$ is a component of the stress tensor. The function $f(x)$ is the conformal map from the euclidean spacetime, with a cut along the interval $A$ and two boundaries, into an annulus of width $2\log L/\epsilon$ and height $2\pi$ \cite{ct}. 
As shown in Appendix \ref{app:app1},  for the ground state of a CFT on the real line, $K_A$ is proportional to the Virasoro generator $L_0$ up to an additive  constant,
\begin{equation}
	K_A=\frac{\pi}{2 \log (L/\epsilon) }\left(L_0-\frac{c}{24} \right),
	\label{bw}
\end{equation}	
therefore
\begin{equation}\label{eq:senzaj}
Z_n=\mathrm{Tr}_Ae^{-2\pi n K_A}=\mathrm{Tr}_A q^{L_0-\frac{c}{24}}, \qquad q=e^{2\pi i \tau}, \tau=\frac{i\pi n}{2\log (L/\epsilon)} 
\end{equation}
with $\epsilon$ the UV cutoff.
This is nothing but a consequence of the celebrated Bisognano-Wichmann theorem \cite{bw-75,bw-76} joined with conformal invariance \cite{ct,ch}.

\subsection{Charged moments and $SU(2)_k$ characters}\label{sec:charged}

We are interested in how $Z_n$ can be resolved in the different $j$ sectors of our theory, $\mathcal{Z}_{n}^{j}(\tau)$. 
The Hilbert space $\mathcal{H}_A$ is a linear combination of the modules, $\mathcal{M}_{j}$, corresponding to a given representation labelled by $j$, 
$\mathcal{H}_A =\oplus_j  n_j\mathcal{M}_{j}$  with coefficients $n_j$.
In order to achieve our goal, let us focus on the charged moments for the conserved quantity $J_0^z$. 
In this section $Z_{n}(\alpha,\tau)$ stands for the charged moments \eqref{eq:firstdef} related to this $U(1)$ charge. They can be written as a linear combination of the unspecialised  characters introduced in Eq. \eqref{eq:unspecialised} \cite{xavier}, i.e. 
\begin{equation}\label{eq:conjalpha}
Z_{n}(\alpha,\tau)=\sum_jn_j\chi_{j} (\alpha,\tau),\qquad \chi_{j} (\alpha,\tau) =\mathrm{Tr}_{\mathcal{M}_{j}} q^{L_0-\frac{c}{24}}e^{i \alpha J_0^z}.
\end{equation}
Here the trace is over all states in the representation with highest weight $j$ and level $k$, which belong to the module $\mathcal{M}_{j}$. 
The index $j=0,\frac{1}{2} \dots \frac{k}{2}$ labels all the unitary representations of the Kac-Moody algebra of $SU(2)_k$ \cite{difrancesco}. 
At this point we are forced to make some physical assumptions on the allowed values of $n_j$, i.e. on the structure of the entanglement spectrum of the CFT.
For example the approach  reviewed in Section \ref{sec:su2g} comes from the continuum limit of an $SU(2)$ spin chain.
In that case, the total spin of the subsystem $A$ would be either integer or half-integer depending on the parity of the length $L$ of the subsystem.
Consequently, $n_j=0$ for half-integer $j$ when $L$ is even, while $n_j=0$ for integer $j$ if $L$ is odd. 
From a CFT perspective, we conclude that the continuum limit of the spin chain induced boundary conditions at the two entangling point of the subsystem $A$ 
that select only integer or half-integers values of $j$'s (although this is difficult to prove directly, see e.g. \cite{act-17} for a similar issue).
We stress that, from the CFT side, other choices of $n_j$ are also fully legitimate; 
anticipating the result,  they all lead to the same double logarithmic factor in the symmetry-resolved entropies \eqref{eq:SREE_final},  
but to a different $O(1)$ term which is affected by a boundary factor $\log g$ ($g$ is the Affleck-Ludwig non-integer ground state degeneracy \cite{al-91}) resulting 
from the induced boundary CFT at the entangling points; 
in turn this is very similar to what is known for the total entanglement entropy \cite{act-17,cs-17}.
\\ 
Going back to our main computation, the $SU(2)_k$ characters are known in the literature and in order to write them down in a compact form, we first define the level-$k$ theta functions 
\begin{equation}\label{eq:theta}
\Theta_m^{(k)}(\alpha,\tau) \equiv \sum_{n \in \mathbb{Z}+ \frac{m}{2k}}q^{kn^2} y^{kn}, \qquad y=e^{i \alpha}.
\end{equation}
Then the $SU(2)_k$ characters read \cite{difrancesco}
\begin{multline}
\label{eq:characsu2}
\chi_{j}(\alpha,\tau)=\dfrac{\Theta_{2j+1}^{(k+2)}(\alpha,\tau)-\Theta_{-2j-1}^{(k+2)}(\alpha,\tau)}{\Theta_{1}^{(2)}(\alpha,\tau)-\Theta_{-1}^{(2)}(\alpha,\tau)}=\\
\dfrac{\Theta_{2j+1}^{(k+2)}(\alpha,\tau)-\Theta_{-2j-1}^{(k+2)}(\alpha,\tau)}{q^{\frac{1}{8}}(y^{\frac{1}{2}}-y^{-\frac{1}{2}})\prod_{n=1}^{\infty}(1-q^n)(1-y q^n)(1-y^{-1}q^n)}.
\end{multline}
We refer to Appendix \ref{app:app1} for a detailed expansion of this quantity.
\\ In the limit $L \gg \epsilon$, one has $q\simeq 1$, so that a large number of terms contribute to Eq. (\ref{eq:characsu2}). However, using the modular transformation $\tau \to -1/\tau$, we get
\begin{equation}
\chi_{j}(\alpha,\tau)=e^{-\frac{\pi i k (\alpha/(2\pi))^2 }{2\tau}}\sum_{j'}S_{j j'}\chi_{j'}\left( \frac{\alpha}{\tau},-\frac{1}{\tau}\right), \qquad  S_{j j'}=\sqrt{\frac{2}{k+2}}\sin \frac{\pi(2j+1)(2j'+1)}{k+2}.
\end{equation}
In the limit $L \gg \epsilon$, only the term with $j'=0$ survives in the sum, so for $\alpha$ around $0$ the final result is
\begin{equation}
\label{eq:due}
Z_{n}(\alpha,\tau)= \sum_j n_j \chi_{j}(\alpha,\tau)\simeq c_{n,\alpha} \Big(\sum_jS_{j0}n_j\Big) e^{2\log (L/\epsilon)\left[\frac{c}{12}\left(\frac{1}{n}\right)-\frac{k}{2n}\frac{\alpha^2}{4\pi^2}\right]},
\end{equation}
as already obtained in \cite{xavier}. 
Similar techniques have been employed for the entanglement entropies in Ref. \cite{ccp-10}. Here $c_{n,\alpha}$ is a non-universal constant which depends on the cutoff (see also the discussion after Eq. \eqref{eq:chargedmom} for the abelian case).

We already mentioned in Section \ref{sec:2} that we are ultimately interested in the saddle point evaluation of the integral which leads to the evaluation of the symmetry-resolved entropies. As we explain in Sec.~\ref{subsec:su2}, in this case the saddle points are determined by the behaviour of the charged moments around $\alpha=0$ and $\alpha=2\pi$, which correspond to the two elements of the center of $SU(2)$, $Z=\{1,-1 \}$ (see the parameterisation (\ref{eq:paramsu2}) of the elements of $SU(2)$). It is not sufficient to know the asymptotic behaviour (\ref{eq:due}) around $\alpha=0$, we also need to know the one around $\alpha=2\pi$. However, when we set $\alpha' \equiv 2\pi - \alpha $, we observe that
\begin{equation}\label{eq:2pi}
Z_n(\alpha,\tau)=\sum_jn_j\chi_j(2\pi-\alpha',\tau)=\sum_jn_j\mathrm{Tr}_{\mathcal{M}_{j}} (-1)^{2J_0^z}q^{L_0-\frac{c}{24}}e^{i \alpha' J_0^z}=\sum_j n_j (-1)^{2j}\chi_j(\alpha',\tau).
\end{equation}
Since either all $j$ are integer or they are all half-integer, the factor $(-1)^{2j}$ simply reduces to an overall factor $1$ or $-1$, respectively. Moreover, the previous asymptotic expansion yields, for $\alpha'$ around $0$,
\begin{equation}
\label{eq:duepi}
Z_{n}(2\pi-\alpha',\tau)\simeq c_{n,{2\pi-\alpha'}} \Big(\sum_j (-1)^{2j}S_{j0}n_j\Big) e^{2\log (L/\epsilon)\left[\frac{c}{12}\left(\frac{1}{n}\right)-\frac{k}{2n}\frac{\alpha'^2}{4\pi^2}\right]} .
\end{equation}
 To summarize, Eq. \eqref{eq:due} gives the large $L$ behaviour when $\alpha \in [0,\pi]$, while Eq. \eqref{eq:2pi} gives the leading contribution when $\alpha \in (\pi, 2\pi]$. We now turn to the analysis of the integral over all group elements parameterized by (\ref{eq:paramsu2}), where we use these asymptotic behaviours.

\subsection{Projecting the charged moments on the spin $j$ representation}\label{sub:saddle}
\label{subsec:su2}
The idea to project the charged moments on the spin $j$ representation is to use the orthonormality of the group characters with respect to the Haar measure to isolate the contribution from all states of spin $j$ in the trace (\ref{eq:senzaj}), corresponding to the term proportional to the group character $\chi_j(\alpha)$. 
This is done by using the orthonormality of group characters with respect to the Haar measure, i.e.  using the following relation between the matrix representation of the group element $g$ in $\mathcal{H}_A$, $U^A_g$, and the projector $\Pi_j$ on all states transforming in the
represention $j$:
\begin{equation}\label{eq:projector}
\Pi_j=\frac{(2j+1)}{{\rm Vol}(SU(2))}\int_Gd\mu(g) \chi^*_j(g)U^A_g,
\end{equation}
where the factor $(2j+1)$ is the dimension of the representation. Let us observe that if we were interested in a symmetry decomposition of entanglement with respect to both $j,j^z$ (as done in \cite{goldstein}), because of the multiplet structure of $SU(2)$, the factor $2j+1$ in Eq. \eqref{eq:projector} should be removed. In other words, all the $2j+1$ states belonging to the same irrep $j$ give the same contribution to the entanglement.
For $SU(2)$, the group characters are given by
\begin{equation}
\label{eq:characsu2finite}
\chi_{j}(\alpha)=\sum_{m=-j}^{j} y^m=\dfrac{y^{j+\frac{1}{2}}-y^{-j-\frac{1}{2}}}{y^{\frac{1}{2}}-y^{-\frac{1}{2}}} =\frac{\sin ((j+\frac{1}{2})\alpha)}{\sin \frac{\alpha}{2}},
\end{equation}
whose behaviour around $\alpha=0$ is $\chi_j(0)=2j+1$, while around $\alpha=2\pi$ is $\chi_j(2\pi)=(-1)^{2j}(2j+1)$.

As already discussed in section \ref{sub:3.1}, the simplest way to measure invariantly the volume of a group, $SU(2)$ in this case, is to start from the Killing metric in the Lie algebra. 
We can write down a generic element of $SU(2)$ in its exponential form as
\begin{multline}
	\label{eq:paramsu2}
g(x,y,z)=e^{i(x\sigma_x+y\sigma_y+z\sigma_z)/2}= \mathbbm{1}\cos \frac{\alpha}{2}+i\frac{\sin (\alpha/2)}{\alpha}(x\sigma_x+y\sigma_y+z\sigma_z), \\ \alpha=\sqrt{x^2+y^2+z^2} \in [0,2\pi]
\end{multline}
where $(x,y,z)$ are the coordinates of the Lie algebra $\mathfrak{su}(2)$ and $\sigma_i$ the Pauli matrices.
Let us observe that for $\alpha=2\pi$, $g=-\mathbbm{1}$, i.e.  the behaviour of $g$ around $\alpha=0,2\pi$ corresponds to the behaviour around the two elements of the center of $SU(2)$, $\mathbb{Z}_2$, i.e. respectively $+\mathbbm{1}$ and $-\mathbbm{1}$. The Killing form is given by 
\begin{equation}
K(\sigma_i,\sigma_j)=\frac{1}{4}\mathrm{Tr}(\mathrm{ad}\frac{\sigma_i}{2}\, \mathrm{ad}\frac{\sigma_j}{2})=\frac{1}{4}\mathrm{Tr}(\sigma_i\sigma_j)=\frac{\delta_{ij}}{2},\qquad i,j \in \{x,y,z\}
\end{equation}
where we used that $\textsl{g}=2$ for $SU(2)$. Once we have found the Killing form, using Eq. \eqref{eq:Kab}, we can fix the metric $K_{ab}(x,y,z), a,b \in \{x,y,z\}$ and compute 
\begin{equation}\label{eq:volumeformsu2}
\sqrt{\det K_{ab}(x,y,z)}=\sqrt{2}\Big(\frac{\sin (\alpha/2)}{\alpha} \Big)^2.
\end{equation}
We can rewrite it in polar coordinates $(x,y,z)=(\alpha \cos \gamma \sin \beta,\alpha \sin \gamma \sin \beta,\alpha  \cos \beta)$ such that
\begin{equation}\label{eq:volumesu2}
\begin{split}
\mathrm{Vol}(SU(2))=&\sqrt{2}\int_{\sqrt{x^2+y^2+z^2}\leq 2\pi}dx\,dy\,dz\,\Big(\frac{\sin (\alpha/2)}{\alpha} \Big)^2=\\
=&\sqrt{2}\int_0^{2\pi}\int_0^{\pi}\int_0^{2\pi}d\alpha d\beta d\gamma \alpha^2\sin \beta \frac{\sin^2(\alpha/2)}{\alpha^2}=2^{5/2}\pi^2.
\end{split}
\end{equation}
From the volume form in Eq. \eqref{eq:volumeformsu2} we can explicitly write down the Haar measure for $SU(2)$ as 
\begin{equation}\label{eq:volume}
\begin{split}
d\mu(\alpha,\beta,\gamma)&= \sqrt{2}\sin^2 \frac{\alpha}{2} \sin \beta \, d\alpha\,d\beta\,d\gamma \qquad 0\leq\alpha \leq 2\pi, 0 \leq \beta \leq \pi, 0 \leq \gamma \leq 2\pi.\\
\end{split}
\end{equation}
One can also explicitly check that using Eq. \eqref{eq:volume} and the $SU(2)$ characters in Eq. \eqref{eq:characsu2finite}, the orthogonality relation in Eq. \eqref{eq:orthogonality} is satisfied.

Using Eq. \eqref{eq:due}, we get
\begin{multline}
\label{eq:F}
\mathcal{Z}_{n}^{j}(\tau)=\frac{2j+1}{\mathrm{Vol}(SU(2))}\displaystyle \int d\mu(\alpha,\beta,\gamma)\frac{Z_{n}(\alpha,\tau)}{Z_1^n}\chi_j(\alpha)\,\\ \simeq \, \frac{e^{-\frac{nj^2\pi^2}{k\log(L/\epsilon)}}}{2\mathrm{Vol}(SU(2))}(2j+1)^2\Big[\frac{Z_n(0,L)}{Z_1^n} \sqrt{\frac{2^5n^3\pi^9}{k^3\log^3  (L/\epsilon)}}+\frac{Z_n(2\pi,L)}{Z_1^n}(-1)^{2j} \sqrt{\frac{2^5n^3\pi^9}{k^3\log^3  (L/\epsilon)}}\Big]
\\\simeq \, \frac{e^{-\frac{nj^2\pi^2}{k\log(L/\epsilon)}}}{\mathrm{Vol}(SU(2))}\frac{Z_n(0,L)}{Z_1^n} (2j+1)^2\sqrt{\frac{2^5n^3\pi^9}{k^3\log^3  (L/\epsilon)}},
\end{multline} 
where we approximate the first line by two integrals, one around $\alpha=0$, the other around $\alpha=2\pi$. Indeed, for $\alpha \in [0,2\pi]$  there is a saddle point at  $\alpha=0$ and one at $\alpha=2\pi$. The first one corresponds to study the charged moments around $g=\mathbbm{1}$, while the second one around $g=-\mathbbm{1}$, which are the two elements of the center of $SU(2)$. %
Let us stress again that since $j$ is fixed to be integer or half-integer, the factors $1$ or $-1$ overall simplify. 
Eq. \eqref{eq:F} coincides with the result found in Eq. \eqref{eq:FG} once we use Eq. \eqref{eq:volumesu2}.
Also here we have kept the Gaussian factor to get a normalised probability, i.e., 
\begin{equation}\label{eq:normalis}
\sum_j\mathcal{Z}_{1}^{j}(\tau)\simeq \sqrt{\frac{\pi^5}{k^3\log^3  (L/\epsilon)}}\int_0^{\infty}dje^{-\frac{j^2\pi^2}{k\log(L/\epsilon)}} (2j+1)^2=1,
\end{equation} 
where, in the large $L$ limit, we can approximate the sum over the irreducible representation as an integral. \\

As a byproduct of our results, from Eq. \eqref{eq:F}, we can compute the number entropy entanglement entropy, i.e.
\begin{multline}\label{eq:ne}
S^{{\rm n}}=-\sum_j \mathcal{Z}_{1}^{j}(\tau)\log \mathcal{Z}_{1}^{j}(\tau)\simeq -\int_0^{\infty}dj \mathcal{Z}_{1}^{j}(\tau)\log \mathcal{Z}_{1}^{j}(\tau) \\ \simeq \frac{1}{2}\log (k \log L) - 2+\gamma_E -\frac{1}{2}\log \pi   + \frac{3}{2} ,
\end{multline}
with $\gamma_E$ the Euler constant. In full analogy with the $U(1)$ case, see e.g. \cite{riccarda}, 
the leading term of the number entropy is a double logarithm in $L$. The prefactor is $1/2$, exactly like for $U(1)$, but this will not be true in general, as shown in the next section.  
When computing the total entropy, this double log cancels with the same contributions coming  from the configurational entanglement entropy, as we will show in the next paragraph.
\subsection{Result for the symmetry-resolved entanglement}
At this point we can plug the result found in Eq. \eqref{eq:F} into the definition of symmetry-resolved entanglement in Eq. \eqref{eq:RSREE}, i.e. 
\begin{equation}\label{eq:sresu21}
S_n^{j}(L)=\frac{1}{1-n}\log \frac{\mathcal{Z}_{n}^{j}(\tau)}{\mathcal{Z}_{1}^{j}(\tau)^n} \propto \frac{1}{1-n}\log \left[\frac{Z_n(0,L)}{Z_1^n} n^{3/2} \left( (2j+1)^2\frac{1}{2}\sqrt{\frac{\pi^5}{k^3\log^3  (L/\epsilon)}}\right)^{1-n} \right].
\end{equation}
The first ratio in \eqref{eq:sresu21} just gives the total R\'enyi entropy of order $n$ while the other term is
\begin{multline}
\log \left( \frac{1}{2}(2j+1)^2\sqrt{\frac{\pi^5}{k^3\log^3  (L/\epsilon)}}\right)+\frac{3}{2(1-n)}\log (n) = \frac{3}{2}\log (\log L)\\ +\log \left( \frac{1}{2}(2j+1)^2\sqrt{\frac{\pi^5}{k^3}}\right)+\frac{3}{2(1-n)}\log (n)+o(L^0) ,
\label{su222}
\end{multline} 
where we have neglected the (subleading) contributions due to the cutoff $\epsilon$ (see Ref. \cite{riccarda} for the $U(1)$ case in which the contribution 
$O(1/\log L)$ are taken into account, too).
Putting everything together, the symmetry-resolved entropies in the $j$ sector are given by
\begin{multline}\label{eq:sresu2}
S_n^{j}(L)= S_n(L)-\frac{3}{2}\log (\log L)-\frac{3}{2}\log (k)+2\log(2j+1)+\frac{3}{2(1-n)}\log (n)+\\\frac{5}{2}\log( \pi)+o(L^0) .
\end{multline}
Summing up the weighted symmetry-resolved contributions \eqref{eq:sresu2} at $n=1$, we get the configurational entanglement entropy
\begin{equation}\label{eq:ce}
S^{\rm c}=\sum_j \mathcal{Z}_{1}^{j}(\tau)S_1^{j}(L)\simeq S_1-\frac{1}{2}\log (k\log L)+2-\gamma_E-\frac{3}{2}+\\\frac{1}{2}\log( \pi).
\end{equation}
Notice that the prefactor $1/2$ of the double logarithmic term comes from the combination of $3/2$ present already in Eq.~\eqref{eq:sresu2} and another double log coming from the integral of the $\log(2j+1)$ term always in Eq. \eqref{eq:sresu2}.
As an important final sanity check, combining Eqs. \eqref{eq:ne} and \eqref{eq:ce}, we straightforwardly verify that Eq. \eqref{eq:dec} is satisfied and the double logarithmic terms exactly cancel in order to recover the total entanglement entropy, $S_1$.

Eq. \eqref{eq:sresu2} is equivalent to \eqref{eq:car1} and so it just represents a consistency check with some known results. 
However,  this calculation is the starting point to study the entanglement resolution for a WZW-model with an arbitrary symmetry group 
and it would have been difficult to motivate many of some intermediate steps (e.g. the choice of $n_j$, the equivalent saddles from the 
elements of the center, etc.) without having in mind a concrete example.

\section{Symmetry resolution for a general group $G$}\label{sec:5}
This section contains the main results of the manuscript: after some explicit examples of symmetry decomposition for WZW-models, we find a general expression for the symmetry-resolved entanglement of these theories by emphasising its universal features. Since we rely on what has been shown in detail for our case study $SU(2)$ in the previous section, it is important to read it before embarking in the study of this one.

\subsection{Derivation of the main result}

We generalise the method exploited for $SU(2)_k$, using the tools of Section~\ref{sec:3}.

\paragraph{Entanglement Hamiltonian and charged moments}

The entanglement Hamiltonian is still given by Eq. \eqref{bw}, i.e.
\begin{equation}
	K_A=\frac{\pi}{2 \log (L/\epsilon) }\left(L_0-\frac{c}{24} \right),
\end{equation}	
therefore
\begin{equation}
Z_n=\mathrm{Tr}_Ae^{-2\pi n K_A}=\mathrm{Tr}_A q^{L_0-\frac{c}{24}}, \qquad q=e^{2\pi i \tau}, \tau=\frac{i\pi n}{2\log (L/\epsilon)} 
\end{equation}
with $\epsilon$ the UV cutoff. Given an element of the algebra $X\in  \mathfrak{g}$, we define the charged moments as $Z_n(X,\tau)= Z_1^n \mathrm{Tr} [ e^{i X}    \rho_A^n ] $. 
The total Hilbert space decomposes as
\begin{equation}
\mathcal{H}_A=\oplus_r n_r \mathcal{M}_r,
\end{equation}
where $n_r$ gives the multiplicity of the module $ \mathcal{M}_r$ over the Kac-Moody algebra.
Hence, the charged moments $ Z_n(X,\tau)$ can be written as a linear combination of the unspecialised characters  in Eq. \eqref{eq:unspecialised}
\begin{equation}\label{eq:ref}
 Z_n(X,\tau)=\sum_r n_r \chi_r(X,\tau),
\end{equation}
with the same coefficients $n_r$.

\paragraph{Asymptotics of the charged moments.}
In the limit $L \gg \epsilon$, we can use the expansion of the unspecialised characters reviewed in Section~\ref{sec:3} to 
find the large-$L$ asymptotics of the charged moments
\begin{equation}\label{eq:asymptotics2}
	 Z_n(X,L)  \underset{L \rightarrow \infty}{\simeq}      Z_n(0,L)  e^{- \frac{k}{2\pi^2 n}\,K(X,X) \, \log (L/\epsilon)}, \qquad {\rm for} \quad X \in \mathfrak{g} ,
\end{equation}
which is valid until $g=e^{iX}$ is in some small neighborhood of the unit element. However, as we have learnt from the $SU(2)$ case, when we project the charged moments onto the irreducible representations, we have to consider the contributions coming from all the saddle points. Apart from the unit element, the other obvious saddle points correspond to the elements $h \in Z(G)$ of the center of the Lie group. Indeed, as we explain below, the contribution around each element $h$ is proportional to Eq. \eqref{eq:asymptotics2}, up to a constant phase.

Furthermore, for simplicity we will assume that these are the only saddle points contributing to the integral. This seems like a reasonable assumption in view of the $SU(2)$ case, but we do not how to prove that it holds for a general group $G$. It is under this assumption that we arrive at our main result (\ref{eq:SREE_final}). We stress that, even if other saddle points were present, the leading orders in Eq.~(\ref{eq:SREE_final}) would remain unchanged; only the order $O(L^0)$ term in Eq.~(\ref{eq:SREE_final}) would be affected.
 
 If we consider group elements of the form $g = he^{iX}$, we have a slightly different asymptotic behaviour with respect to Eq. \eqref{eq:asymptotics2} when $h$ is not simply the unit element. The unitary matrix $U_g^A$ can be decomposed as $U^A_g = U^A_hU_{e^{
iX}} $, where $U_h^A$ is a representation of $Z(G)$, which is a finite abelian subgroup. Let $\Pi_m^{Z(G)}$,
$m = 1, \cdots, |Z(G)|$, denote the projector onto states in $\mathcal{H}_A$ that transform in the $m$-th irreducible (one-dimensional) representation of $Z(G)$. Irreps of $Z(G)$ are just phases times the identity in each block, i.e.
\begin{equation}\label{eq:centrerep}
U_h=\sum_m e^{i\varphi_m(h)}\Pi_m^{Z(G)}.
\end{equation}
This definition shows that the elements of the center of a group play the same role up to a constant (in $L$) phase. We stress again that our intuition about these multiple saddle points has been suggested by the explicit computations done for the $SU(2)$ case in Section \ref{sub:saddle}.
In our case study $SU(2)$, the center is given by $\mathbb{Z}_2$, and we have already seen that around $g=\mathbbm{1}$, $e^{i\varphi_j(1)}=1$, while around $g=-\mathbbm{1}$, $e^{i\varphi_j(-1)}=(-1)^{2j}$. However, these phases are fixed to be $+ 1$ or $-1$ by the boundary conditions at the entangling points of the subsystem through the coefficients $n_r$ appearing in Eq. \eqref{eq:ref}. 
In a similar way, for the general case we assume that all the non-zero $n_r$ in Eq. \eqref{eq:ref} correspond to representations $r$ that are in the same block in Eq. \eqref{eq:centrerep}, i.e. there is a single term in the sum \eqref{eq:centrerep}. As a consequence, $U_h$ is fixed to be simply a phase, $U_h=e^{i\varphi(h)}$, and the asymptotic expression in the neighborhood of $h$ reads
\begin{equation}\label{eq:asymptoticsh}
	 Z_n(X,L)  \underset{L \rightarrow \infty}{\simeq}      Z_n(0,L)  e^{i\varphi(h)}e^{- \frac{k}{2\pi^2 n}\,K(X,X) \, \log (L/\epsilon)}, \qquad {\rm for} \quad X \in \mathfrak{g}, \, h \in Z(G) .
\end{equation}
\paragraph{Projecting the charged moment onto the representation $r$ using the orthonormality of group characters.}

The main idea of our approach is to use the orthonormality of group characters to extract the contribution of the representation $r$ from $ \mathrm{Tr} [  g  \rho_A^n ]$, i.e. recalling $\mathcal{Z}_n^r (L) = \mathrm{Tr} [  \Pi_r  \rho_A^n ]$,
\begin{eqnarray}
	\mathcal{Z}_n^r (L) &=& \frac{{\rm dim}(r)}{{\rm Vol}(G)} \int d\mu (g)  \mathrm{Tr} [  g  \rho_A^n ]    \chi_{r}^*(g)  .
\end{eqnarray}
Similarly to what has been discussed after Eq. \eqref{eq:projector}, if we were interested in a symmetry resolution involving also the quantum numbers labelling the states within an irreducible representation $r$ (e.g. $j^z$ for the case study $SU(2)$), the prefactor ${\rm dim}(r)$ should be removed. The reason is that the entanglement Hamiltonian $K_A$ is independent of them because it commutes with the corresponding charge generators, therefore each quantum number in a given irrep $r$ gives the same contribution to $ \mathrm{Tr} [  g  \rho_A^n ] $, from which the prefactor ${\rm dim}(r)$ arises.
Strictly speaking, $ \mathrm{Tr} [  g  \rho_A^n ]$ differs from the charged moments built with $\mathrm{Tr} [ h e^{i X}    \rho_A^n ]$ because  
the former is valid for arbitrary group elements $g \in G$, while the latter is valid only for elements in a neighbourhood of the element $h$ in the center of the group.
%, as obtained with the exponential map $X \in \mathfrak{g} \mapsto e^{i X} \in G$. 
Nevertheless, the knowledge of only  $\mathrm{Tr} [ h e^{i X}    \rho_A^n ]$ is enough for our aims because we are going to use a saddle point integral that is dominated by the 
elements of the group in some neighbourhood of $h \in Z(G)$.\\
%Around the unit element of $G$, i
Around $h$, it is convenient to use the local coordinate chart ${\rm x} \mapsto g({\rm x}) = h e^{i \sum_a x_a J^a}$. Replacing the integral over the whole group $G$ by the integral over the neighbourhood of  $h$ parametrised  by this chart, we have
\begin{eqnarray}
	\mathcal{Z}_n^r (L) &=& \sum_{h \in Z(G)}\frac{{\rm dim}(r)}{Z_1^n{\rm Vol}(G)} \int d\mu (he^{i \sum_a x_a J^a})   Z_n( {\bf x},L)  \chi_{r}^*(he^{i \sum_a x_a J^a}) \nonumber \\
	&=& \sum_{h \in Z(G)} \frac{{\rm dim}(r)}{Z_1^n{\rm Vol}(G)} \int \sqrt{{\rm det} K ({\bf x}) } d{\bf x}   Z_n({\bf x},L)    \chi_{r}^*(he^{i \sum_a x_a J^a})   .
\end{eqnarray}
Now we use the asymptotics (\ref{eq:asymptoticsh}), and then we do a saddle point approximation around the elements of the center of the group,
\begin{eqnarray}\label{eq:final}
\nonumber	&\mathcal{Z}_n^r (L) &\\ 
\nonumber &\underset{L \rightarrow \infty}{\simeq}&   \sum_{h \in Z(G)} \frac{Z_n(0,L)}{Z_1^n}  \frac{{\rm dim}(r)e^{i\varphi(h)}}{{\rm Vol}(G)} \int \sqrt{{\rm det} K({\bf x}) }   e^{- \frac{k}{2\pi^2 n} \, \sum_{a,b}  x_a x_b K(J^a,J^b) \, \log (L/\epsilon)}   \chi_{r}^*(he^{i \sum_a x_a J^a})  ~d{\bf x}  \\
\nonumber &\simeq & \sum_{h \in Z(G)}  \frac{Z_n(0,L)}{Z_1^n}  \frac{{\rm dim}(r)e^{i\varphi(h)}}{{\rm Vol}(G)}  \left( \frac{ 2 \pi^3 n}{k \, \ln (L/\epsilon)} \right)^{ {\rm dim} (G)/2} \mathrm{Tr}  \left[ h e^{-\frac{\pi^2n}{2k\log (L/\epsilon)}\sum_{a,b} K^{-1}(J^a,J^b)J^aJ^b}\right] ,\\
 &\simeq &  \frac{Z_n(0,L)}{Z_1^n}  \frac{|Z(G)|}{{\rm Vol}(G)}  \left( \frac{ 2 \pi^3 n}{k \, \ln (L/\epsilon)} \right)^{ {\rm dim} (G)/2}  {\rm dim}^2 (r)  e^{-\frac{\pi^2n}{k\log (L/\epsilon)}C^{(2)}_r } ,
%	&\simeq &   \frac{Z_n(0,L)}{Z_1^n}  \frac{1}{{\rm Vol}(G)}  \left( \frac{ 2 \pi^3 n}{k \, \log (L/\epsilon)} \right)^{ {\rm dim} (G)/2}  {\rm dim} (r),
\end{eqnarray}
where in the last step we used that $(K(J^a,J^b)=\delta_{ab}/2)$, 
\begin{multline}
\mathrm{Tr} \, h e^{-\frac{\pi^2n}{2k\log (L/\epsilon)}\sum_{a,b} K^{-1}(J^a,J^b)J^aJ^b}=e^{-i\varphi(h)}\mathrm{Tr}\,e^{-\frac{\pi^2n}{k\log (L/\epsilon)}\sum_{a} J_aJ^a}\, \\ =e^{-i\varphi(h)} e^{-\frac{\pi^2n}{k\log (L/\epsilon)}C^{(2)}_r}\mathrm{Tr}\, \mathbbm{1}_{{\rm dim}(r)\times {\rm dim}(r)}=e^{-i\varphi(h)}e^{-\frac{\pi^2n}{k\log (L/\epsilon)}C^{(2)}_r}{\rm dim}(r),
\end{multline}
and $C^{(2)}_r$ labels the eigenvalues of the quadratic Casimir operator of $G$. We also remark that the evaluation of the Gaussian integral holds for $J^a=O(\sqrt{\log L})$, such that the saddle-point approximation is valid.  \\
From Eq. \eqref{eq:final} for $n=1$, we also read that the probability introduced in Eq. \eqref{eq:probab} is in the large $L$ limit
\begin{equation}\label{eq:probab11}
p^{(r)}(L)\simeq \frac{|Z(G)|}{\mathrm{Vol}(G)}\left( \frac{ 2 \pi^3 }{k \, \log (L/\epsilon)} \right)^{ {\rm dim} (G)/2}  {\rm dim}^2 (r)e^{-\frac{\pi^2}{k\log (L/\epsilon)}C^{(2)}_r }.
\end{equation}
Interestingly, the normalisation of that probability distribution, $\sum_r p^{(r)}(L) = 1$, in the large $L$ limit, leads us to the following asymptotic formula relating the quadratic Casimir operator and the dimension of irreducible representations of $G$, 
\begin{equation}
	\label{eq:conjecture} 
\lim_{\eta \rightarrow 0^+} \left( 2 \pi  \eta \right)^{ {\rm dim} (G)/2}  \sum_r  {\rm dim}^2 (r)e^{- \eta C^{(2)}_r } \, = \, \mathrm{Vol}(G) .
\end{equation}
Here the sum is over all irreps $r$ of $G$. 
[This is the reason why the factor $|Z(G)|$ has dropped. It reenters if one restricts the sum to irreps $r$ that transform identically under the action of the center $Z(G)$, see also the discussion in Section \ref{sec:subsuN}.] 
This formula may be viewed as an analog of the one for finite groups, that says that the square of dimensions of all irreps is equal to the order of the group, since one may regard ${\rm Vol}(G)/(2 \pi \eta)^{{\rm dim}(G)/2}$ as the order of some finite approximation of the continuous Lie group $G$. 
 
Unfortunately, we have not been able to find formula (\ref{eq:conjecture}) in the mathematics literature. 
It is very likely that it comes from results on the Plancherel formula for Lie groups (see, e.g., Refs. \cite{harish1954plancherel,vergne1982poisson}), but we have not been able to find it in the explicit form (\ref{eq:conjecture}) (also Ref. \cite{fegan} is closely related to this subject). 
Nevertheless, in the following section we will check its validity explicitly for the group $SU(N)$, for some values of $N$, using the actual form of the quadratic Casimir operator.

\paragraph{Final result.}

Finally, the symmetry-resolved entanglement entropy is
\begin{eqnarray}
\nonumber	S_n^r(L) \, = \, \frac{1}{1-n} \log  \frac{\mathcal{Z}_n^r(L) }{ (  \mathcal{Z}_1^r(L)   )^n }  ,
\end{eqnarray}
leading to our final result
\begin{multline}\label{eq:SREE_final2}
 S_n^r (L)   =   S_n(L) -  \frac{{\rm dim} (G)}{2} \log (\log L) + 2\log {\rm dim}(r)  - \log   {\rm Vol}(G)  +\log |Z(G)| \\ + \frac{{\rm dim} (G)}{2} \left( - \log k + \frac{\log n}{1-n} + \log( 2\pi^3 ) \right)  .
\end{multline}

This is the main result of this work: at leading order, the symmetry-resolved entanglement entropy satisfies equipartition, i.e. it is equally distributed in the different symmetry sectors. 
Interestingly, we find the term $2\log (\mathrm{dim}(r)) $, at ${O}(L^0)$ which explicitly depends on the specific representation of the group $G$, breaking equipartition. 
This is different from what was found in the literature for the abelian case, where the first terms breaking equipartition usually occur at order $O((\log L)^{-2})$ 
(the two results are compatible since in the abelian case $\dim(r)=1$ always). 
Also the prefactor of the double logarithmic correction has a universal behaviour which depends on the dimension of the group. 
Actually, the entire form \eqref{eq:SREE_final2} at order $O(L^0)$ is universal since the ultraviolet cutoff is fully encoded in the total entropy.

\subsection{The explicit example of $SU(N)$}\label{sec:subsuN}

As an example, let us specialize to the case of the WZW model based on the group $G=SU(N)$. 
In order to use Eq. \eqref{eq:SREE_final2} we should provide the values for $\dim (G)$, $|Z(G)|$, ${\rm Vol}(G)$, and $\dim{r}$. 
As well known, the dimension of $SU(N)$ is just 
\begin{equation}
\dim(SU(N))= N^2-1.
\label{dimsun}
\end{equation}
and the center of the group is $Z(SU(N)) = \mathbb{Z}/(N \mathbb{Z})$, with order $|Z(SU(N))|=N$. The dual Coxeter number is $\textsl{g} = N$, so the Killing form, normalized as in Eq.~(\ref{eq:killing}), is $K(X,Y) = {\rm Tr} (X^\dagger Y) $. The invariant metric on the group is then
\begin{equation}
d\mu(g)=\mathrm{Tr}(ig^{-1}\partial_a g, i g^{-1}\partial_b g).
\end{equation} 
For $SU(N)$ we can write the equivalence
\begin{equation}\label{eq:iso}
\frac{SU(N)}{SU(N-1)}=S^{2N-1},
\end{equation}
 where $S^{2N-1}$ is the sphere of unit radius embedded in $\mathbb{R}^{2N}$. 
Around the identity $\mathbbm{1}_N$, we can relate the local coordinates of the sphere and $SU(N)$ through  
\begin{equation}
g=\mathbbm{1}_n+iy_N\mathrm{diag}\{\frac1{N-1}, \cdots ,\frac1{N-1},1 \}+\sum_{j=1}^{N-1}(z_i e_{jN}-\bar{z}_je_{Nj})
\end{equation} 
where  $\sum_{j=1}^N|z_j|^2=1, z_N=x_N+iy_N$ and $e_{kj}$ is a matrix with 0 everywhere except for a single 1 at position $(k,j)$. Therefore, the metric around the unit element of $SU(N)$ reads
\begin{equation}
d\mu(g)=\frac{N}{N-1}dy^2_N+2\sum_{j=1}^{N-1}[dx^2_j+dy^2_j],
\end{equation}
and using the equivalence in Eq. \eqref{eq:iso} we find
\begin{equation}
\mathrm{Vol}(SU(N))=\sqrt{\frac{N}{N-1}}2^{N-1}\mathrm{Vol}(SU(N-1))\mathrm{Vol}(S^{2N-1}), \quad \mathrm{Vol}(S^{2N-1})=\frac{2\pi^N}{(N-1)!}.
\end{equation}
By induction, the final formula for $SU(N)$ turns out to be
\begin{equation}
\mathrm{Vol}(SU(N))=\sqrt{\frac{N}{N-1}\frac{N-1}{N-2}\dots 2}\prod_{j=1}^{N-1}\frac{(2\pi)^{j+1}}{j!}=\sqrt{N} \frac{(2\pi)^{\frac{N^2+N}{2}-1}}{G(N+1)},
\label{volfin}
\end{equation}
with $G(N+1)=\prod_{j=1}^{N-1}{j!}$ the Barnes G function.
\\ Another ingredient we need is the dimension of the representation $r$, denoted by $\mathrm{dim}(r)_{SU(N)}$, where now $r=(\lambda_1,\lambda_2, \dots, \lambda_{N-1})$ is a set of integers that univocally identifies the irreducible representations. Its explicit expression can be found in \cite{difrancesco}
\begin{equation}
\mathrm{dim}(r)_{SU(N)}=\frac{\displaystyle\prod_{l=1}^{N-1}\prod_{k=l}^{N-1}\Big(\sum_{m=k-l+1}^k \lambda_m +l\Big)}{\prod_{k=1}^{N-1}k!}.
\label{dimrsun}
\end{equation} 
Plugging Eqs. \eqref{dimsun},  \eqref{volfin}, and \eqref{dimrsun} into \eqref{eq:SREE_final2} we 
get our final form for the symmetry-resolved R\'enyi entropies in $SU(N)_k$ WZW models. \\
\begin{table}[t]
%\bigskip
\begin{center}
\begin{tabular}{||p{0.3cm} ||p{1.1cm}||p{4.6cm}||p{4.6cm}||p{0.9cm}||p{1.1cm}||}
\hline
  $N$&   ${\rm Vol}(G)$& \quad \qquad  $C^{(2)}_{r=(\lambda,\cdots,\lambda_{N-1})}$  & \quad \qquad   ${\rm dim}(r)$  &   $|Z(G)|$ &   ${\rm dim}(G)$ \\
\hline
\hline
  2 & $2^{5/2}\pi^2$  & $\frac{1}{4}\lambda_1(\lambda_1+2)$ &$\lambda_1+1$ & 2 & 3\\
 \hline
 3 & $\sqrt{3}2^4\pi^5 $  & $\frac{1}{3}(\lambda^2_1+\lambda^2_2+\lambda_1\lambda_2+3(\lambda_1+\lambda_2))$ &$\frac{1}{2}(\lambda_1+1)(\lambda_2+2)(\lambda_1+\lambda_2+2)$ & 3 & 8\\
\hline
 4 & $\frac{2^{8}}{3}\pi^9$  & $\frac{1}{8}(3\lambda^2_1+4\lambda^2_2+3\lambda^2_3+4\lambda_1\lambda_2+2\lambda_1\lambda_3+4\lambda_2
 \lambda_3+12\lambda_1+6\lambda_2+12\lambda_3)$ &$\frac{1}{12}(\lambda_1+1)(\lambda_2+1)(\lambda_3+1)(\lambda_1+\lambda_2+2)(\lambda_2+\lambda_3+2)(\lambda_1+\lambda_2+\lambda_3+3)$ & 4 & 15\\
\hline
\end{tabular}
\renewcommand{\arraystretch}{1}
\caption{The table summarises all the information required for computing the probabilities in Eq. \eqref{eq:probab11} for $G=SU(2), SU(3), SU(4)$.}\label{fig:table}
\end{center}
\end{table}
Finally, it is worth elaborating on the allowed values of $\lambda_j$'s that in turn affect also the normalisation of the probability in Eq. \eqref{eq:probab11}. 
Let us consider explicitly the groups $G=SU(2),SU(3),SU(4)$ for which  all the needed ingredients are summarised in Table \ref{fig:table}.\\
For large $\lambda_i$'s, we can drop all the subleading terms in the expression of the Casimir eigenvalues and ${\rm dim}(r)$, so that the probabilities extracted from ${\cal Z}_1(r)$ read
\begin{equation}\label{eq:su234}
\begin{split}
p_{\lambda_1}^{SU(2)} & \simeq\pi^{5/2}\left( \frac{ 1 }{k \, \log (L/\epsilon)} \right)^{ 3/2} \lambda_1^2e^{-\frac{\pi^2}{4 k\log (L/\epsilon)}\lambda_1^2}, \\
p^{SU(3)}_{\lambda_1,\lambda_2}&\simeq\sqrt{3}\pi^7\left( \frac{  1 }{k \, \log (L/\epsilon)} \right)^{ 4}\frac{1}{4}(\lambda_1\lambda_2)^2(\lambda_1+\lambda_2)^2e^{-\frac{\pi^2}{3 k\log (L/\epsilon)}(\lambda^2_1+\lambda^2_2+\lambda_1\lambda_2)},\\
p^{SU(4)}_{\lambda_1,\lambda_2,\lambda_3}&\simeq\pi^{27/2}\left( \frac{  1 }{k \, \log (L/\epsilon)} \right)^{ 15/2}\frac{1}{12\sqrt{2}}(\lambda_1\lambda_2\lambda_3)^2(\lambda_1+\lambda_2)^2(\lambda_2+\lambda_3)^2(\lambda_1+\lambda_2+\lambda_3)^2\\
&e ^{-\frac{\pi^2}{8 k\log (L/\epsilon)}(3\lambda^2_1+4\lambda^2_2+3\lambda^2_3+4\lambda_1\lambda_2+2\lambda_1\lambda_3+4\lambda_2\lambda_3)}.
\end{split}
\end{equation}
They do not contain yet information about the possible values of $\lambda_j$'s, that are still all the integers. However, the boundary conditions we imposed through the coefficients $n_r$  in Eq. \eqref{eq:ref} induce some constraints on the possible values that $\lambda_i$'s can assume. 
For example, for $SU(2)$, we identified $\lambda_1=2j$: since $j$ can be either integer or half integer, $\lambda_1$ can be only even or odd, respectively. 
This implies that when we sum over the possible representations, we should consider only the values of $\lambda_i$'s which are compatible with our boundary conditions encoded in \eqref{eq:ref}. 
Since in the large $L$ limit, it holds $\sum_r p_r \to \int dr p(r)$ we can take into account the possible values of $\lambda_i$ simply by   
multiplying the final result by the ratio $1/|Z(G)|$ between the allowed cases over the total ones, i.e. the correct continuum limit in $r$ is 
$\sum_r p_r \to \frac1{|Z(G)|} \int dr p(r)$. 
Accordingly, the probabilities in Eq. \eqref{eq:su234} satisfy
\begin{equation}
\begin{split}
&\frac{1}{2}\int  d\lambda_1\, p_{\lambda_1}^{SU(2)}=1, \\
&\frac{1}{3}\int  d\lambda_1d\lambda_2\, p_{\lambda_1,\lambda_2}^{SU(3)}=1, \\
&\frac{1}{4}\int  d\lambda_1d\lambda_2d\lambda_3\, p_{\lambda_1,\lambda_2,\lambda_3}^{SU(4)}=1.
\end{split}
\end{equation}
If we set $\lambda_1=2j$, we get the computation for the $SU(2)$ case in Eq. \eqref{eq:normalis}.\\
After this digression, we can determine the number entropy \eqref{eq:dec} for $SU(N)$ as
\begin{equation}
S^{{\rm n}}=\frac{N-1}{2}\log(k\log L)+\frac{N^2-1}{2}-\frac{N^2-1}{2}\log(2\pi^3)+\log \frac{{\rm Vol}(SU(N))}{|Z(SU(N))|}+N(N-1)\log \pi-c,
\label{nnn}
\end{equation}
%where $c=\int dr \frac{2}{\mathrm{Vol}(SU(N))}(  2 \pi^3 )^{(N^2-1)/2}  {\rm dim}^2 (r)e^{-\pi^2C^{(2)}_r }\log  {\rm dim} (r)$.
where $c=G(N+1)2^{(3-N)/2}\frac{(\pi)^{-\frac{1+N}{2}+N^2}}{\sqrt{N}}\int dr   {\rm dim}^2 (r)e^{-\pi^2C^{(2)}_r }\log  {\rm dim} (r)$. In order to derive this result for $SU(N)$, we used that 
\begin{equation}
\frac{2}{|Z(G)|} \int dr p(r) \log(\mathrm{dim}(r))=\frac{N(N-1)}{2}\log \frac{k\log L}{\pi^2}+c.
\end{equation}
Also in the general $SU(N)$ case, the leading term of the number entropy is a double logarithm with a prefactor depending explicitly on $N$.
Finally, for $SU(N)$ we can also compute the configurational entropy in Eq. \eqref{eq:dec}, which reads 
\begin{equation}
S^{{\rm c}}=S_1-\frac{N-1}{2}\log(k\log L)-\frac{N^2-1}{2}+\frac{N^2-1}{2}\log(2\pi^3)-\log \frac{{\rm Vol}(SU(N))}{|Z(SU(N))|}-N(N-1)\log \pi+c,
\label{nnc}
\end{equation}
Combining Eqs. \eqref{nnn} and \eqref{nnc} specialised to $SU(N)$, it is possible to show that Eq. \eqref{eq:dec} is satisfied; 
in particular that the double logarithmic terms in the symmetry-resolved entanglement and in the number entropy cancel each other in the sum.

\section{Conclusions}\label{sec:concl}
In this manuscript, we considered the decomposition of the entanglement entropy into the various sector of a non-abelian symmetry.  
In particular, we studied the resolution of the entanglement entropy in WZW-models, which are associated to a group $G$ and their symmetry algebra is a Kac-Moody algebra.
Writing the charged moments as a linear combination of the unspecialised characters of these theories and using their modular properties, we have computed the resolved partition functions, i.e. the ones which take into account the preserved symmetry by including only states in a given representation of the group. 
We first characterised the general scaling behaviour of the charged moments;  
then we focused on the integration measure over the group manifold and the group characters around the elements of its center to extract the symmetry-resolved moments and R\'enyi entropies.
Our physically more relevant findings are: (i) symmetry-resolved entanglement satisfies equipartition at leading order;
(ii) this equipartition is broken at $O(L^0)$ by a term depending only on the dimension of the irrep; 
(iii) the coefficient in front of the double logarithmic correction to the R\'enyi entropies is universal and it is equal to half of the dimension 
of the symmetry group of the model;
(iv) the difference $S_n-S_n^r$, between symmetry-resolved R\'enyi entropy and total one, is universal up to order $O(L^0)$ and the cutoff enters only 
in higher order terms; (v) as a byproduct of the aforementioned results, we can find the expression of the number entropy for $SU(N)$ up to $O(L^0)$ and show that 
Eq. \eqref{eq:dec} is indeed satisfied.

It is worth mentioning that while throughout all the manuscript we only wrote the results for the ground state of a single interval in the infinite line, 
it is easy
%by properly modifying the parameter $\log (L/\epsilon)$ in Eq. \eqref{eq:senzaj}, 
to generalise our findings to different situations such as a finite interval in an infinite system at finite temperature, or finite interval in a finite 
system by using standard conformal transformations on the worldsheet.
Moreover, it is also possible to adapt our results to the massive field theories, obtained by adding a relevant perturbation to the critical ones, in the regime $L\gg m^{-1}$
using the ideas of Ref. \cite{mdgc-20}.

Finally, our findings also lead to few very natural questions and generalisations. 
The most natural one is how other entanglement measures decompose in the sectors of a non-abelian symmetry and if there is some
important difference with the abelian case \cite{goldstein1,mbc-21,c-21,cc-21}. 
A second one is whether it is possible to generalise the form factor bootstrap program of Refs. \cite{hcc-21,dhc-20,hcc-21b} to the resolution of non-abelian symmetries. 
A last one is to identify the holographic dual of the symmetry-resolved entanglement entropy for theories with non-abelian symmetry and compare it with our results, 
as already done for the abelian case in \cite{znm-20}.

\section*{Acknowledgments}
We thank Paola Ruggiero and Benoit Estienne for useful discussions. PC and SM acknowledge support from ERC under Consolidator grant number 771536 (NEMO).  JD acknowledges support from CNRS International Emerging Actions under the grant QuDOD.

\appendix
\section*{Appendices}
\section{Entanglement Hamiltonian} \label{app:app1}
In this appendix, we try to sketch the main steps which lead to the identification of the entanglement Hamiltonian with the Virasoro generator, as written in Eq. \eqref{eq:senzaj}. Moreover, we give an alternative brief argument about the entanglement equipartition for the $U(1)$ compact boson.\\
Consider the ground state of a 1+1 dimensional CFT Hamiltonian $H =\int_{-\infty}^{\infty} dx h(x)$. 
To cure the ultraviolet divergences in the construction of the entanglement Hamiltonian of the interval $A=[-L/2,L/2]$ and  it is custom to remove the degrees of freedom in 
a small circle in the Euclidean space-time around the entangling points at $\pm L/2$, see e.g. \cite{ct};
on these circle specific boundary condition are imposed and affect the  entanglement spectrum of the interval $A$ \cite{ct,act-17}. 

Under these circumstances, the entanglement Hamiltonian $K_A$ is expressed as an integral of the Hamiltonian density $h(x)$ \cite{ch,ct}
\begin{equation}\label{eq:Ka}
K_A=\int_{-L/2+\epsilon}^{L/2-\epsilon} dx \frac{L^2/4-x^2}{L} h(x)+{\rm const}
\end{equation} 
where the constant enforces $\mathrm{tr} (\rho_A) = 1$. 
$K_A$ is easily rewritten in terms of the Virasoro generator $L_0$, i.e.  %can be expressed in the complex coordinates $z,\bar{z}$ as 
\begin{equation}
L_0=\frac{1}{2\pi i} \left(\int_{\mathcal{C}} dz z^{n+1} T(z)-\int_{\mathcal{C}} d\bar{z}\bar{z}^{n+1} T(\bar{z}) \right), 
\end{equation}
where the integration contour $\mathcal{C}$ is a semicircle going counterclockwise around the origin. 
Using standard conformal mappings, we get $L_0$ in the cut plane of Eq.  \eqref{eq:Ka}, obtaining \cite{hrv-20}
\begin{equation}
L_0=\frac{2\log( L/\epsilon)}{\pi}\int_{-L/2+\epsilon}^{L/2-\epsilon} dx \frac{L^2/4-x^2}{L} h(x)+\frac{c}{24}\left(1+\frac{4\log (L/\epsilon)}{\pi^2} \right).
\label{A3}
\end{equation}
Comparing Eqs. \eqref{eq:Ka} and \eqref{A3}, we find that $K_A$ is proportional to $L_0$ up to an additive constant:
\begin{equation}\label{eq:bw1}
K_A=\frac{\pi}{2\log (L/\epsilon)}\left(L_0-\frac{c}{24}\right)+{\rm const}.
\end{equation}
The result in Eq. \eqref{eq:bw1} and the CFT structure of a compact free boson can be combined to easily show the equipartition of entanglement for a system endowed with a $U(1)$ symmetry without computing the charged moments as done in the main text. 
In terms of the conformal modes $a_n$, the Virasoro generator $L_0$ can be written as 
\begin{equation}
L_0=\sum_{n>0}a_{-n}a_n+\frac{1}{2}a_0^2. 
\end{equation}
The term $\frac{a_0^2}{2}$ commutes with $L_0$ and is the generator of translations in the target space of the compact bosonic field, i.e. the generator of the $U(1)$ symmetry
of interest.  
Moreover, the Hilbert space factorises into a tensor product of $(U(1)\, \mathrm{charges}\,) \otimes \, (\mathrm{Fock \, space})$. 
Using this form of $L_0$ in the entanglement Hamiltonian \eqref{eq:bw1}, the symmetry-resolved entanglement hamiltonian is just the piece corresponding to a given 
eigenvalues $q$ of $a_0^2$. Hence the distribution is Gaussian in $q$ and equipartition for large $L$ follows from central limit theorem.   
\\ %This has been dubbed equipartition of the entanglement entropy, as mentioned above. \\
It is worth mentioning why these ideas do not generalise to non abelian symmetries. Taking as a case study $SU(N)$, 
the splitting of the entanglement Hamiltonian in two pieces still occurs. As described in Eq. \eqref{eq:virsug}, one could use the Sugawara construction to show that it splits into \cite{difrancesco}
\begin{equation}
L_0=\frac{1}{2(k+\textsl{g})}\sum_a(J_0^{a \, 2}+2\sum_{n>0}J^a_{-n}J^a_n),
\end{equation}
i.e. the quadratic Casimir operator and the other modes of the current. 
The Casimir operator commutes with $L_0$, but a factorisation of the Hilbert space similar to the $U(1)$ case does not hold as it can be shown for $SU(2)$.
Indeed, taking the $SU(2)_k$ characters in Eq. (\ref{eq:characsu2}), and expanding them with the help of Eq. \eqref{eq:characsu2finite}, we get
\begin{equation}\label{eq:char15}
\dfrac{\Theta_{2j+1}^{(k+2)}(\alpha,\tau)-\Theta_{-2j-1}^{(k+2)}(\alpha,\tau)}{q^{\frac{1}{8}}(y^{\frac{1}{2}}-y^{-\frac{1}{2}})} \simeq q^{-\frac{k}{8(k+2)}+\frac{j(j+1)}{k+2}}\left(\chi_{j}(\alpha) -q^{k+1-2j}\chi_{k+1-j}(\alpha) +\dots \right).
\end{equation}
Here, we recognise  the prefactor as $q^{h-\frac{c}{24}}$, where the central charge is  $c=\frac{3k}{k+2}$ and the conformal weight of the ground state is $h=\frac{j(j+1)}{k+2}$. However, it is evident from the right hand side that a factorisation between the $SU(2)$ charges and the currents modes $J^a_n$ does not hold.
%, otherwise we should have something like $q^{h-\frac{c}{24}}\chi^{SU(2)}_{\ell}$. 

\section{$N$-component free non-compact boson with $O(N)$ symmetry}\label{app:app2}
In this section we consider a conformal field theory which has an $O(N)$ symmetry but which is not a WZW-model. 
It is an $O(N)$ non compact free boson, defined by the action 
\begin{equation}
S=\frac{1}{2}\int d^2x \sum_{i=1}^N (\partial \phi_i)^2.
\end{equation}
\\ 
First, let us recall what happens in the abelian case $N=2$. 
The action is invariant under linear transformations acting upon the vector $\vec{\phi}=(\phi_1,\phi_2)$ and preserving the norm $\phi_1^2+\phi_2^2$, 
i.e. a rotation in the target space. To this rotation we can associate a single complex number $e^{i\alpha}$.
The conserved charge restricted to the interval $A$  is given by
\begin{equation}
Q_A=\int_A dx( (\partial \phi_1) \phi_2-\phi_1(\partial \phi_2)).
\end{equation}
The charged moments have been computed in Ref. \cite{mdgc-20}, getting
\begin{equation}\label{eq:o2}
\mathrm{Tr}(\rho_A^ne^{i\alpha Q_A}) \propto e^{\log L\left( \frac{1}{6}(\frac{1}{n}-n)+\frac{\alpha^2}{4\pi^2 n}-\frac{|\alpha|}{2\pi n}\right)},
\end{equation}
whose Fourier transform is given by ($q$ labels the eigenvalues of $Q_A$)
\begin{equation}\label{eq:bosonic}
\begin{split}
\mathcal{Z}^q_n(L)&=\int_{-\pi}^{\pi}\frac{d\alpha}{2\pi} e^{iq\alpha}\mathrm{Tr}(\rho_A^ne^{i\alpha Q_A})\propto \frac{Z_n(0,L) n}{\log L},\\
S^q_n(L)&\sim S_n-\log (\log L)+\frac{1}{1-n}\log (n),
\end{split}
\end{equation}
which is different from the results of $U(1)$ compact boson in \eqref{eq:intro}, although the global symmetry is the same.\\
%which is invariant under linear transformations acting upon the vector $\vec{\phi}=(\phi_1,\phi_2,\dots \phi_N)$ and leaving invariant its norm 
%\begin{equation}
%|\vec{\phi}|^2=\sum_i^N \phi_i \phi_i.
%\end{equation}
%This job is done by orthogonal transformations, i.e. matrices forming the rotation group in $N$ dimensions, $O(N)$.
Let us now move to the general $O(N)$ case. We label by $f_{abc}$ the structure constants of the Lie group $O(N)$ and we define the matrices $[T^a]_{bc}=-if_{abc}$,  
satisfying $[T^a,T^b]=if_{abc}T^c$. The $N(N-1)/2$ conserved currents are 
\begin{equation}
J^{a}_{\mu}=-i\partial_{\mu}\phi_{i}T_{ij}^{a}\phi_j,
\end{equation}
and the corresponding conserved charges are
\begin{equation}
Q_A^{a}=\int_A dx (\partial \phi_i)T^{a}_{ij}\phi_j.
\end{equation}
At this point we rely on a power counting argument similar to what we did in Eq. \eqref{eq:final} where each integral was Gaussian and led to a contribution proportional to $\sqrt{\log L}^{{\rm dim}(G)}$, while here the integral over $O(N)$ leads to a term proportional to $(\log L)^{{\rm dim}(O(N))}$, as shown by Eq. \eqref{eq:bosonic} when $N=2$.
Using the behaviour of the group character of $O(N)$ around the identity, 
we get 
\begin{equation}\label{eq:oN}
{\mathcal{Z}}^r_n(L)\simeq \frac{Z_n(0,L)}{\mathrm{Vol}(O(N))Z_1^n}\mathrm{dim}^2(r) (\log L/n)^{-{\rm dim}(O(N))},\quad {\rm dim}(O(N))=\frac{N(N-1)}{2},
\end{equation}
and so
\begin{multline}
S^{(r)}_n(L)= S_n(L)-{\rm dim}(O(N))\log (\log L) +2\log \mathrm{dim}(r) \\+ {\rm dim}(O(N))\frac{\log (n)}{1-n}-\log \mathrm{Vol}(O(N))+o(L^0).
\end{multline}
Let us stress one main difference with respect to the result found in Eq. \eqref{eq:SREE_final2} for a WZW model with $G=SO(N)_{k=1}$: apart from the $O(L^0)$ terms, the prefactor in front of the double logarithmic correction is ${\rm dim}(O(N))$, not ${\rm dim}(O(N))/2$. 
This result shows that the resolution of the entanglement strictly depends on the model which implements the symmetry we are interested in and not only on the symmetry itself: WZW-models with $SO(N)_{k=1}$ symmetry corresponds to $N$ real independent free fermions while in this appendix we are working with $N$ real independent free bosons.

\end{document}